\documentclass[runningheads,a4paper]{llncs}

\input{psfig.sty}
\usepackage{amssymb}
\usepackage{graphicx}
\usepackage{caption}

\usepackage{url}
\urldef{\mailsa}\path|{alfred.hofmann, ursula.barth, ingrid.haas, frank.holzwarth,|
\urldef{\mailsb}\path|anna.kramer, leonie.kunz, christine.reiss, nicole.sator,|
\urldef{\mailsc}\path|erika.siebert-cole, peter.strasser, lncs}@springer.com|    
\newcommand{\keywords}[1]{\par\addvspace\baselineskip
\noindent\keywordname\enspace\ignorespaces#1}

\usepackage[table,dvipsnames]{xcolor}
\usepackage{colortbl}
\usepackage{listings}
\usepackage{multirow}
\usepackage{amsfonts}
\usepackage{amsmath}
\usepackage{algpseudocode}

\usepackage{algorithm}

\usepackage{syntax}
\usepackage{mathtools}
\usepackage{algorithmwh}
\usepackage{multicol}
\usepackage{float}
\usepackage{semantic}
\usepackage{algorithm}
\usepackage{listings}
\usepackage{caption}
\usepackage{framed}
\usepackage{graphics}
\usepackage{tikz}
\usepackage{pgf}
\usetikzlibrary{positioning,decorations.pathreplacing,arrows,automata}
\usepackage[latin1]{inputenc}
\usepackage{courier}
\usepackage{array}
\floatstyle{boxed}

\DeclareMathAlphabet{\mathpzc}{OT1}{pzc}{m}{it}

\makeatletter
\def\BState{\State\hskip-\ALG@thistlm}
\makeatother

\newcolumntype{P}[1]{>{\centering\arraybackslash}p{#1}}

\newsavebox\lstbox
\begin{document}

\mainmatter  

\title{Presburger-Definable Parameterized Typestates}

\titlerunning{Parameterized Typestates}

%
%
 \author{Ashish Mishra \and Deepak D'souza \and Y. N. Srikant}
%
%
 \authorrunning{A. Mishra, D. D'souza, Y. N. Srikant}
\authorrunning{}
%
\institute{Indian Institute of Science, Bangalore}
%
%

\maketitle
\begin{abstract}
Typestates are good at capturing dynamic states of
a program as compared to normal types that can capture static structural
properties of data and program. Although useful, typestates are suitable
only for specifying and verifying program properties defined using finite-state abstractions. Many useful dynamic properties of programs are not finite-state definable. To address these issues,
we introduce parameterized typestates (p-typestates). p-typestates associate
a logical property with each state of regular typestate, thereby allowing specification 
of properties beyond finite-state abstractions. We present a dependent type system 
to express and verify p-typestate properties and a typestate-oriented core 
programming language incorporating these dependent types. Automatic inductive 
type-checking of p-typestate properties usually requires a programmer to provide
loop invariants as annotations. Here we propose a way to calculate loop invariants automatically, using loop acceleration techniques for Presburger definable transition systems.
\end{abstract}

\section{Introduction}
\label{introduction}

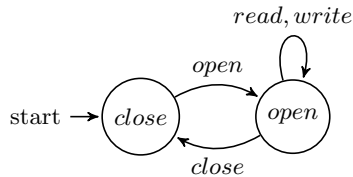
\begin{figure*}

\centering
\begin{tikzpicture}[->,>=stealth',shorten >=1pt,auto,node distance=2cm,
                    semithick]
  \tikzstyle{every state}=[fill=white,draw=black,text=black]

  \node[state,initial] (B)                    {$close$};
  \node[state] 		     (A)  [right of=B]      {$open$};

  \path (A) edge [loop above] node {$read, write$} (A)
            edge [bend left]	       node {$close$} (B)
    
	 (B) edge [bend left] node {$open$} (A);

\end{tikzpicture}
\caption{Typestate property for FileManager}
\label{fig:simplets}
\end{figure*}

Typestates~\cite{typestate,typestates-for-objects,Plaid,TSOP,Fink} are important programming language concept useful in enforcing protocols and properties over data, software components or whole programs. They form a component of general Behavioral types~\cite{behavioral-types-seminar} capturing the dynamic state of a program as compared to the normal types which capture the static structure of data and programs. This aids in making software reliable and robust by early elimination of many semantic errors in programs by checking  correct implementation and usage of a protocol. 
A classic example of typestate is a \textsf{FileManager} which allows a \textsf{File} object to have a set of operations defined over it, \textit{viz.} \textsf{open, close, read} and \textsf{write}. A subset of these operations are valid on a \textsf{File} object based on its current state.  Figure~\ref{fig:simplets}, shows a finite automaton representing valid operations in $open$ and $closed$ states of a \textsf{File} object and pre- and postconditions for each method.

Typestate-oriented programming languages~\cite{TSOP,Plaid} introduce typestates into programming languages as first class members. It allows a programmer to explicitly encode her intent in a program (which would otherwise be documented as a class document) and allows implementation of typestate abstractions and properties directly in the program which may then be enforced statically or dynamically using a suitable type system. 
Although efficient at capturing properties related to the state of data, the original typestates can only express program properties from the regular language domain~\cite{typestates-for-objects,typestate}. Consider a commonly occurring non-regular property defined by the language of matching parenthesis, its variants are quite prevalent in various domains, for example, a well formed \textit{XML} document must have a matching string of opening and closing \textit{XML} tags. Since the property belongs to \textit{context-free languages}, it cannot be captured using regular typestates. This limitation of regular typestate impedes the effectiveness of various typestate analyses, type systems for typestate verification and typestate oriented programming languages.

In the lack of a more expressive typestate, such properties are checked using costly, error prone and difficult to debug runtime checks~\cite{JML}, or are expressed using expressive but undecidable type systems~\cite{ESC,Coq,Agda}, or some are converted to abstract models and then verified using existing non-regular verification techniques~\cite{Flat-acceleration-symbolic}. These abstract model based verification do not allow correct-by-construction practical implementation of real systems.  

The main idea of our work is to remove this expressive limitation of regular typestates.
We present a generalized notion of typestates called  Parameterized typestates (p-typestate). Parameterized typestates allow associating a \textit{Presburger logical formula} to the states of a regular typestate, this allows expressing richer typestate properties. \textit{Presburger logic} is the First Order Logic of integers with ``+''(addition). Thus, a p-typestate state is a pair ($\phi, s$), where $\phi$ is a Presburger definable formula over set of auxiliary variables and $s$ is a state of a regular typestate. We present a dependent type system to specify and verify p-typestate properties and present a typestate-oriented core programming language incorporating these dependent types. Further, automatic inductive type checking p-typestate properties usually requires a programmer to annotate loop invariants, to placate this burden from the programmer, we present a novel loop invariant calculation approach using loop acceleration technique for Presburger definable transition system.

The major contributions of our work are listed below-
\begin{itemize}
\item We introduce \textit{Parameterized typestates}.
\item We present a dependent type system to capture and enforce parameterized typestate properties and define a core typestate-oriented language with static p-typestate checking.
\item We present a novel loop-invariant calculation technique based on acceleration techniques for Presburger definable systems.
\item We further show the effectiveness of the p-typestates by implementing many real world programs enforcing p-typestate properties which cannot be specified using regular typestates.
\end{itemize}

\section{Overview}
\label{overview}

Consider Figure~\ref{fig:SimpleXMLParser}, a variant of XMLParser for Java, over a binary set of $XML$ elements, $\Sigma$ =  \{ $\langle el \rangle, \langle /el \rangle$ \}. A string $z \in$  $\Sigma^{*}$ is \textit{well-formed} iff, $\forall z = xy$, $D(x)$ $\geq$ 0 and $D(z)$ = 0, where $D(p)$ = (\# $\langle el \rangle$ in $p$ - \# $\langle /el \rangle$ in $p$) is called the \textit{Dyck distance} of a string $p$. 

The \textsf{XMLParserSimple} uses runtime checks to enforce this well-formedness property of an XML document. The property can be formally defined by a \textit{p-typestate property automaton} (formally defined later) of Figure~\ref{p-typestate-property-automata}. Counters \textsf{ns} and \textsf{ne} count the number of start and end elements scanned. The boolean field \textsf{isOpen} stores the \textit{open} or \textit{close} states of the parser. Method \textsf{startReading} starts reading the input document changing the state of the parser from \textit{close} to \textit{open} and is valid only in the \textit{close} state of the parser. Methods \textsf{scanStartElement} and \textsf{scanEndElement} scan a start and end element respectively and are valid in the \textit{open} state of the parser iff the substring $z$ scanned thus far has \textit{Dyck dictance} greater than or equal to zero, i.e. \textsf{ns} $\geq$ \textsf{ne}. Finally method \textsf{endOfFile} checks the terminal condition for well-formedness of $z$, i.e. \textsf{ns} $==$ \textsf{ne}. 
The reader can see that regular typestates can specify and statically verify the runtime checks for method \textsf{startReading} since the property being verified can be defined using a finite-state abstraction. However such a check for \textsf{scanStartElement}, \textsf{scanEndElement} and \textsf
{endOfFile} requires checking value of counters \textsf{ns} and \textsf{ne} and thus cannot be specified using regular typestates.

Parameterized typestates extend regular typestates to handle such non-regular properties by associating extra information with its states. Figure~\ref{fig:brtypestatexmlparser} presents a program using parameterized typestates to statically verify the property enforced dynamically in Figure~\ref{fig:SimpleXMLParser}.

The program uses the syntax and attributes common to a typestate-oriented programming language like Plaid~\cite{Plaid} with p-typestate annotations of pre- and post p-typestate conditions. Later in Section~\ref{corelanguage} we will introduce a p-typestate-oriented core language to make these examples precise. The state \textsf{XMLSimpleParser} defines an $XML$ parser as a state with defined p-typestate property, methods and fields. The type definition at line 3 defines a new parameterized typestate, parameterized with the number of $\langle el \rangle$ scanned (\textsf{ns}), the number of $\langle /el \rangle$ scanned (\textsf{ne}), a logical formula defining a property over these variables and a state (\textsf{Parser}) from the regular typestate of a Parser, belonging to the set \{ Open, Close, $\perp$ \}. Each method is annotated with \textit{contracts} defining pre- and postconditions. Using these annotations we can mechanically verify the correctness of the implementation of these methods, i.e. each implementation assumes the precondition and guarantees the annotated postcondition. This is termed as \textit{protocol implementation verification}. Using these verified implementations of methods and expressions, we can inductively verify a p-typestate property of the program in a modular way using Floyd-Hoare style assertion reasoning. A p-typestate property verification checks that for each execution of the program, the sequence of method invocations is a legal sequence permitted by the p-typestate property automaton. This is termed as \textit{protocol usage verification}.
\begin{figure*}[htbp]
\begin{lstlisting}[xleftmargin=1pt, basicstyle=\selectfont\ttfamily\scriptsize, numbers=left, extendedchars=true, numberstyle=\tiny, xrightmargin=1pt, mathescape=true]
public class XMLParserSimple{
	private int ns; /* number of startElement */
	private int ne; /* number of endElement */
	List<Element> elementsScanned = new ArrayList<Element>();
	private boolean isReading = false;
	private boolean isOpen = false;
	public void startReading(){
		if(isOpen $\mid \mid$ isReading){reportFatalError("IllegalStateException");}
		else{ this.isOpen = true; this.isReading = true;	}
	}
	public Element scanStartElement(Element el){
		String elementName = element.getElementName();
		if(!isOpen)
			reportFatalError("IllegalStateException");
		else if(ns < ne)
			reportFatalError("Element mismatch", new Object[] {elementName});
		else if(!(el instanceof StartElement))
			reportFatalError("Illegal Element Type", new Object[] {elementName});
		else{@ \bf{this}@.ns++;}
	}
	public Element scanEndElement(Element el){
			(...)this.ne++;
	}
	public boolean endOfFile(Element el){
		if(!isOpen){ reportFatalError("IllegalStateException");}
		else if(!(ns == ne)){
		      reportFatalError("Element mismatch" new Object[] {elementName});
		else { this.isReading = false; this.isOpen = false; }
	}
}
\end{lstlisting}
\caption{Code snippet XMLParserSimple with runtime invariant checks}
\label{fig:SimpleXMLParser}
\end{figure*}

The \textsf{main} method uses these states and methods to scan a sequence of $\langle el \rangle$ and $\langle /el \rangle$ via  methods \textsf{scanStartElement} and \textsf{scanEndElement} respectively. The methods are first invoked explicitly for some finite number of times ($b1$ and $b2$ respectively) and then conditionally while iterating a list \textsf{storedElements} of elements. We assume that the size of \textsf{storedElements} is statically known and is represented by $N$. This could be achieved by using \textsf{SizedList}, an implementation of sized version of a List from our language library. Here we use Java \textsf{List} for simplicity. The \textsf{while} removes an \textsf{Element, el} from \textsf{storedElements} and conditionally scans a start ($\langle el \rangle$) or an end ($\langle /el \rangle$) element based on the kind of the element removed. Let us try to verify the correctness (\textit{usage verification}) of the \textsf{main} program for the input p-typestate property counter system (figure ~\ref{p-typestate-property-automata}).
\begin{figure*}[htbp]
\begin{lstlisting}[xleftmargin=1pt, basicstyle=\selectfont\ttfamily\scriptsize, numbers=left, extendedchars=true, numberstyle=\tiny, xrightmargin=1pt, mathescape=true] 
state XMLParserSimple case of Parser{

 	type SafeSimpleParser : Pi (ns, ne, ns $\geq$ ne) $->$ Parser;
	List<Element> elementsScanned = new ArrayList<Element>();
	Boolean isReading;
	method unique Element startReading()[@\bf{unique SafeSimpleParser (n, m) $->$ Close $\gg$ unique SafeSimpleParser (n, m) $->$ Open}@]{
	 	this.isReading = true; 
		this <- uninue SafeSimpleParser(n, m) -> Open;
	}
	method unique Element scanStartElement(unique StartElement selement)[@\bf{unique SafeSimpleParser (n, m, n $\geq$ m) $->$ Open $\gg$ unique SafeSimpleParser (n'=n+1, m'=m, n' $\geq$ m') $->$ Open}@]{
	 	elementsScanned.add(selement);
		this <- uninue SafeSimpleParser(n+1, m) -> Open;
	}
	method unique Element scanEndElement(unique EndElement eelement)[@\bf{unique SafeSimpleParser (n, m, n $\geq$ m) $->$ Open $\gg$ unique SafeSimpleParser (n'=n, m'=m+1, n' $\geq$ m') $->$ Open}@]{
		elementsScanned.add(eelement);
		this <- unique SafeSimpleParser (n, m+1) -> Open;
	}
	method boolean endOfFile()[@\bf{unique SafeSimpleParser (n, m, n == m) $->$ Open $\gg$ unique SafeSimpleParser (m'= m, n'= n, n' == m' ) $->$ Close}@]{
		this.isReading = false;
		this <- unique SafeSimpleParsr(n , m, n == m) -> Close;
	}
}
method void main(){
  var unique SafeSimpleParser (0, 0) $->$ Close safeParser = 
	  new XMLParserSimple{elementsScanned = new ArrayList<Element>(); isReading=flase;};
  List<Element> storedElements = new ArrayList<Element>(); 
  //populate storedElements 
  (...)
  safeParser.startReading();
  @\{safeParser.scanStartElement(new StartElement());\}@ //b1 times   
  
  @\{safeParser.scanEndElement(new EndElement()); \}@ //b2 times
  
  var numberScanned = 1;
  val N = storedElements.size();
  while(numberScanned <= N){
    Element el = storedElements.remove();
    match(el){
      case (StartElement){ 
	scanStartElement(el);
      }
      case (EndElement){
	scanEndElement(el);
      }
      default {} 
     numberScanned++; 
    };
  };
  scanStartElement(new StartElement());

}


\end{lstlisting}
\caption{SimpleXMLParser using p-typestate (p-typestate annotations are shown in \bf{bold})}
\label{fig:brtypestatexmlparser}
\end{figure*}
Starting from the p-typestate definition instantiation at line 25, we create bounded variables (\textsf{ns, ne}) as integer counters to count the number of start and end elements scanned thus far. Each statement checks the precondition and updates the counters and the states according to the postcondition. We model these counter values and updates as Presburger formulas, while states (Open, Close, $\perp$) are expressed explicitly. Readers can observe that the value of counters associated with the parser before entering the \textsf{while} loop is given by the formula $\phi_{in}$ := (\textsf{ns = }$b1$ $\wedge$ \textsf{ne = }$b2$ $\wedge$ \textsf{ns - ne} $\geq$ 0), where $b1$ and $b2$ are the number of explicit calls to \textsf{scanStartElement} and \textsf{scanEndElement} between lines 30 to 32.
Let us assume that the programmer has provided a Presburger formula as an invariant property for the \textsf{while} loop as $\phi_{l}$ := ($\exists k_1, \exists k_2$ \textsf{ns' = ns + }$k_1$ $\wedge$ \textsf{ne' = ne + } $k_2$ $\wedge$  0 $\leq$ $k_1, k_2$ $\leq N$  $\wedge$ $k_1 + k_2 \leq N$ ), where the primed version of variables represent the values post each iteration of the loop and $N$ is the size of the list \textsf{startOrEndElements}. Plugging in the loop invariant in the incoming p-typestate, gives us the possible p-typestate before the call to the final \textsf{scanStartElement} (line 49). We update the p-typestate value according to the precondition of \textsf{scanStartElement} declaration. This gives us the constraint before this call which must be satisfied for call to be valid, $\phi_{pre}$ := (\textsf{ns + }$k_1$ $\geq$ \textsf{ne + }$k_2$ $\wedge$ 0 $\leq$ $k_1, k_2$ $\leq N$ $\wedge$ $k_1 + k_2 \leq N$). We solve this set of Presburger logical formulas by passing them to Z3~\cite{z3} solver for validity checking. Its easy to see in this case that, the formula is valid iff $ b1 - b2 \geq N$ which also defines the correctness condition for the \textsf{main} method. Figure~\ref{fig:main}(a) shows a type correct \textsf{main} for $b1$ = 4, $b2$ = 1 and $N$ = 3 while, Figure~\ref{fig:main}(b) shows a \textsf{main} violating the property with $b1$ = 4, $b2$ = 2 and $N$ = 3. The variable and method names have been abbreviated to save space. 

This example demonstrates three main challenges involved in checking a rich non-regular property using a static typestate checking: 1) Capturing logical formulas or properties associated with a state of a parameterized typestate (Handled in Section~\ref{parameterizedtypestate}). 2) Inductively generating a set of logical formulas for each statement based on its abstract semantics and solving the formula for validity (Handled in Section~\ref{corelanguage}). 3) Finding a precise loop invariant associated with loops, to verify a property, without the assumption of the programmer annotation (Handled in Section~\ref{inference}). Our work allows us to perform all these steps for p-typestate verification automatically.

\begin{figure*}[htbp]
\begin{minipage}{0.5\textwidth}
\begin{lstlisting}[caption={(a)}, xleftmargin=1pt, basicstyle=\selectfont\ttfamily\scriptsize, numbers=left, extendedchars=true, numberstyle=\tiny, xrightmargin=1pt, mathescape=true] 
method void main(){
  (...)
  sP.scanSE(new StartElement());   
  sP.scanSE(new StartElement());
  sP.scanSE(new StartElement());   
  sP.scanSE(new StartElement());      
  
  sP.scanEE(new EndElement()); 
  var numberScanned = 1;
  val N = storedElements.size(); 
  while(numberScanned <= 3){
    Element el = 
	storedElements.remove();
    match(el){
      case (StartElement){ 
	scanStartElement(el);
      }
      case (EndElement){
	scanEndElement(el);
      }
      default {} 
     numberScanned++; 
    };
  };
  scanSE(new StartElement());
}
\end{lstlisting}
\label{fig:main-a}
\end{minipage}
\hspace{0.25cm}
\begin{minipage}{0.5\textwidth}
\begin{lstlisting}[caption={(b)}, xleftmargin=1pt, basicstyle=\selectfont\ttfamily\scriptsize, numbers=left, extendedchars=true, numberstyle=\tiny, xrightmargin=1pt, mathescape=true] 
method void main(){
  (...)
  sP.scanSE(new StartElement());   
  sP.scanSE(new StartElement());
  sP.scanSE(new StartElement());   
  sP.scanSE(new StartElement());       
  sP.scanEE(new EndElement()); 
  sP.scanEE(new EndElement()); 
  var numberScanned = 1;
  val N = storedElements.size(); 
  while(numberScanned <= 3){
    Element el = 
	storedElements.remove();
    match(el){
      case (StartElement){ 
	scanSE(el);
      }
      case (EndElement){
	scanEE(el);
      }
      default {} 
     numberScanned++; 
    };
  };
  scanSE(new StartElement());
}
\end{lstlisting}
\label{fig:main-b}
\end{minipage}
\caption{correct and incorrect \textsf{main} for Figure~\ref{fig:brtypestatexmlparser}
(*sP : safeParser, scanSE : scanStartElement, scanEE : scanEndElement)}
\label{fig:main}
\end{figure*}

\section{Parameterized Typestate}
\label{parameterizedtypestate}
In this section we discuss the main idea of Parameterized typestate and how it solves the problems discussed in Sections~\ref{introduction} and~\ref{overview}. We extend the typestate concept to define Parameterized typestate(p-typestate). 

\subsection{Formal Definitions}
\begin{definition}[Typestate ~\cite{typestate}]
\textnormal{Given a strongly typed language $\mathpzc{L}$ and an extensible set of types $\mathpzc{T}$, the typestate (or regular typestate) associated with a type $\tau \in \mathpzc{T}$ is a \textit
{partially ordered set} $(S(\tau), \prec)$, such that there exists a ubiquitous typestate $\perp$ corresponding to the initial state of a term $t$ of type $\tau$ of the language and $(S(\tau), \prec)$ forms a lower semi-lattice with $\perp$ being the bottom element.
}
\end{definition}
For example, the regular typestate S(\textsf{File}) for Simple \textsf{File} object in Figure~\ref{fig:simplets} is a poset (\{$Open, Close$, $\perp$\}, $\prec$) with an ordering relation $\prec$ = ($\perp \prec Close \prec Open$). 
 
\begin{definition}[Parameterized Typestate]
\textnormal{
The p-typestate $S_{\Psi}(\tau)$ associated with a type $\tau \in \mathpzc{T}$ is a set defined as a Cartesian product $(\Psi \times S(\tau))$, where $\Psi$ is a class of logical formulas over a set of \textit{auxiliary} variables and $S(\tau)$ is a regular typestate set for $\tau$ . The class $\Psi$ acts as a \textit{parameter} to the parameterized typestate. For instance, if $S(File)$ = $\{ \{open, close, \perp\}, \prec \}$ is the typestate for a type $File$ capturing the state of a variable in a strongly typed language, then the p-typestate set is given by $\{ (\phi, s) \mid \phi \in \Psi, s \in \{open, close, \perp\}$\}.} 
\end{definition}
For a strongly typed language $\mathpzc{L}$, we formally define a program $\mathbb{P}$ in $\mathpzc{L}$. A program $\mathbb{P}$ is a sequence of expressions where each expression is defined as a pair $<op, \overline{V}>$ where $op$ is an operation from the set of valid operations $\mathpzc{O}$ and $\overline{V} = <v_1, v_2, ..., v_N>$ is an indexed set of operands to $op$. Each opertaion $op \in \mathpzc{O}$ has a signature $\mathpzc{T}(op)$ = $\langle t_1 : \tau_1, t_2 : \tau_2, ..., t_n : \tau_n \rangle$ specifying the type of its operands and result, with each $\tau_i$, representing the type of the $i_{th}$ operand term $t_i$. Each $v_i$ is an actual argument for the formal operand with the restriction on type of $v_i$ $Typeof(v_i)$ = $\tau_i$.
To extend $\mathpzc{L}$ to include p-typestate each type  $\tau \in \mathpzc{T}$ is mapped to a set of p-typestate defined as a set $S_{\Psi}(\tau)$ and to track p-typestate changes for a type $\tau$ due to an operation, we define p-typestate transitions ($\delta : S_{\Psi}(\tau) \mapsto 2^{S_{\Psi}(\tau)}$) for each operation $op \in \mathpzc{O}$ as ($Pre_{op, i} \gg \{ Post_{op, i, k} \}$), such that 
\begin{itemize}
\item $Pre_{op, i}$, is the p-typestate precondition for the $i_{th}$ operand $v_i$,  $Pre_{op, i} \in \mathpzc{S}_{\Psi}(Typeof(v_i))$, defines the required p-typestate for $v_i$ for $op$ to be applicable.
\item For each different outcome $k$, $k$ = 1, 2, ..., $m$, $Post_{op, i, k} \in \mathpzc{S}_{\Psi}(Typeof(v_i))$ represents the typestate for $v_i$, when $op$ terminates with outcome $k$.
\end{itemize}

Different instances of $\Psi$ in the p-typestate definition gives different families of p-typestates, with varying degree of expressiveness and decision properties. For instance, instantiating $\Psi$ with First Order Logic(FOL) gives a highly expressive specification system like that of ESC/Java, but undecidability of validity checking of FOL makes it ill suited to automatic type-checking or verification. On the other extreme end of the spectrum, instantiating $\Psi$ with a tautological formula ``True'' gives the system of regular typestate with limited expressiveness. The optimal choice for $\Psi$ should be a logical family with considerable expressiveness and decidable and efficient decision properties. The family of Presburger arithmetic logic~\cite{paf} is an appropriate choice for this giving us a Presburger definable typestate.

\begin{definition}[Presburger Definable Typestate]
\textnormal{A Presburger definable typestate is an instance of p-typestate such that, the p-typestate $S_{\Psi}(\tau)$ associated with a type $\tau \in \mathpzc{T}$ is a set defined as a cartesian product $(\Psi_{\mathpzc{P}} \times S(\tau))$, where $\Psi_{\mathpzc{P}}$ is the Presburger arithmetic logical family.
} 
\end{definition}

\textbf{In the rest of this work we focus on this instance of p-typestate leaving the study of other instances for future work}.

\subsubsection{p-typestate Property Automata}

A parameterized typestate property automata is a dynamic system defining p-typestate properties, similar to the way finite automata defines regular typestate properties. As it turns many useful functional correctness properties of interests can be modeled using p-typestate property automata. We provide a formal definition of p-typestate property automata.

\begin{definition}[p-typestate Property Automata]
\textnormal{A p-typestate property automata $PTS$ is defined as a tuple $\langle Q, \Sigma, \Psi, D, \delta \rangle$
where $Q$ is a finite set of states, $\Sigma$ defines a finite set of actions, $\Psi$ is a set of formulas from a logical family, $D$ represents the domain of values for variables in $\Psi$ and $\delta \subseteq 2^{(Q \times \Psi \times D)}$ is the transition relation defining transitions over $\Sigma$.}
\end{definition} 
%

A $PTS$ is a parameterized transition system. Given a p-typestate property automaton $PTS$, each instance of $\langle \Psi, D \rangle$ gives a different Interpretation of $PTS$ with ranging expressiveness and decision properties. For instance, an interpretation $I_1$ := ($\Psi$ = $\mathpzc{P}$ ,Presburger Arithmetic Logic formulas,  D = $\mathbb{Z}$), defines a \textit{Counter systems}, which is in fact a p-typestate property automaton for \textit{presburger definable typestate}. Similarly, $I_2$ := ($\Psi$ = $\mathpzc{B}$, Boolean formulas , D = $\mathbb{N}$), where defines a simple \textit{Minsky machines}~\cite{minsky-machines}.

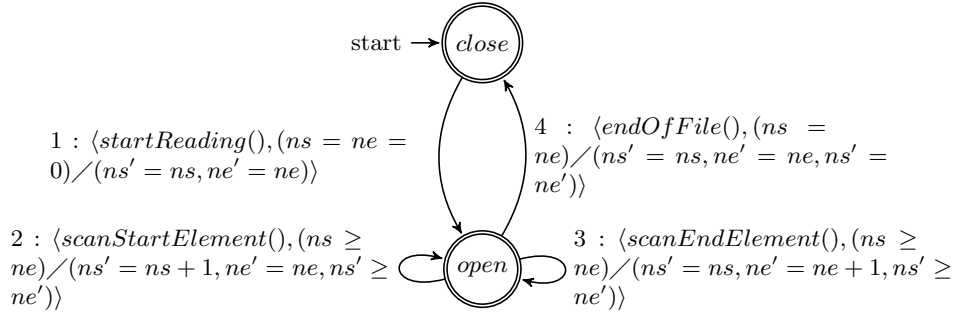
\begin{figure*}[htbp]
\centering
\begin{tikzpicture}[->,>=stealth',shorten >=1pt,auto,node distance=3cm,
                    semithick]
  \tikzstyle{every state}=[fill=white,draw=black,text=black]

  \node[initial,accepting,state] (A)                    {$close$};
  \node[accepting,state,double] (B) [below of=A]        {$open$};

  \path  (A) edge [bend right, text width=5cm, left] node {$ 1: \langle startReading(), (ns=ne=0) \diagup (ns'=ns,ne'=ne) \rangle$} (B)
  
	  (B) edge [loop left, text width=5cm] node {$ 2: \langle scanStartElement(), (ns \geq ne)\diagup (ns'=ns+1, ne'=ne, ns' \geq ne') \rangle$} (B)
	      edge [loop right, text width=5cm] node {$ 3: \langle scanEndElement(), (ns \geq ne)\diagup (ns'=ns,ne'=ne+1,ns' \geq ne') \rangle $} (B)
	      edge [bend right, text width=5cm, right] node {$4:\langle endOfFile(), (ns=ne) \diagup (ns'= ns,ne'=ne,ns'=ne')\rangle $}(A);

\end{tikzpicture}
 \caption{A p-typestate Property Automata for XMLParserSimple (Figure~\ref{fig:brtypestatexmlparser}) }
\label{p-typestate-property-automata}
\end{figure*}

Example -  
Figure ~\ref{p-typestate-property-automata} presents a p-typestate property automata for the property of the SimpleXMLParser example presented earlier. Each transition is labeled with a tuple (a $\in \Sigma$, $\phi \in \Psi$ $\diagup$ $\phi' \in \Psi$), where $\phi$ and $\phi'$ are Presburger formulas over auxiliary variables satisfied by the pre and post states on an action a $\in \Sigma$. We have numbered the transitions for illustrious purposes. For instance, transition 2 represents the p-typestate transition for the \textit{scanStartElement} method call. The variables $ns, ne$ represent the number of $\langle el \rangle$ and $\langle /el \rangle$ scanned. The primed variables represent these same variables in post state. The transition defines $\delta$ for the underlying p-typestate property automata with \textit{Pre} = $\forall ns, ne. ns \geq ne$ while \textit{Post} = $\forall ns, ne. \exists ns', ne'. ns'=ns+1, ne'=ne+1, ns' \geq ne'$ . The transitions does not involve set $D$ of domain values, this domain becomes significant to assign values to various bounded index variables of Presburger formulas while checking a correct usage of a protocol.

\begin{definition}[p-typestate property Verification Problem]
\label{def:pts-verification}
\textnormal{Consider a program $\mathbb{P}$ in a strongly typed programming language $\mathpzc{L}$ (formally defined later), which is a sequence of method definitions annotated with p-typestate contracts and a \textsf{main} method defined as an inter-procedural control flow graph over statements and expression of the language, and a p-typestate property automaton $PTS$ with actions $\Sigma$ defined over language statements and expressions. A p-typestate property verification problem is to check that each inter-procedurally valid path in main is a legal sequence generated by the input p-typestate property automaton.}

\end{definition}

For example, consider Figure~\ref{fig:main}(a), Every possible strings of $\langle el \rangle$ , $\langle /el \rangle$ generated by the \textsf{main} method is guaranteed to be a legal sequence generated by the input p-typestate property automaton. On the other hand, there may exists a valid  string ($\langle el \rangle^4.\langle /el \rangle^5.\langle el \rangle$) generated by the \textsf{main} in Figure~\ref{fig:main}(b), which is not a legal sequence generated by the property automaton.


%

\subsubsection{Capturing p-typestate with Dependent Types}

There are various possible ways to capture and track these p-typestate properties associated with each type $\tau$ in $\mathpzc{T}$. For instance, Java Modeling Language~\cite{JML} semantically tracks such a set using annotations and assertions in First Order Logic which are checked at runtime. Although expressive, this incurs a great runtime cost. 
Most intuitive and natural approach is using the expressive power of dependent type theory. Dependent Types~\cite{Martin-Lof-1972} are types which can depend on the terms of other types. Unfortunately, this expressiveness of general dependent type theories like Martin L$\ddot{o}$f's type theory (theory behind Coq and Agda) comes at the cost of complex type-checking. The type-checking problem is undecidable in general in theses theories. We chose a sub-fragment of these theories which is expressive, and yet has a decidable type-checking. This sub-fragment restricts the index terms of dependent types to Presburger arithmetic logical formulas.

More formally, a p-typestate property associated with a type $\tau$, $S_{\Psi}(\tau)$ = ($\Psi_{\mathpzc{P}} \times S(\tau)$) is  defined as a Dependent function type $\Pi(\Psi_{\mathpzc{P}}, S(\tau)).t$, where $\Psi_{\mathpzc{P}}$ represents Presburger arithmetic logic family. Each instance of this dependent function type, represented by ($\phi$, s).$t$ represents a state of the p-typestate. The exact approach of defining, instantiating and transiting these p-typestates are presented in the core-language section~\ref{corelanguage} of the paper.

\section{Core Language and Type System}
\label{corelanguage}
In this section we present a formal definition of our dependent type-system implementing p-typestate. To formalize the concept we present a core restricted dependently typed language, its syntax, semantics and typing rules defining the p-typesate type system. 

We focus on a language with monomorphic dependent types and static type checking. The language's restricted dependent type makes it expressive enough to define and verify crucial p-typestate program properties. But it is practical as compared to other decidable and less expressive typestate oriented languages~\cite{Plaid,typestates-for-objects} on the one hand and highly expressive languages with undecidable type-checking like Coq or Agda on the other. 

\subsection{Syntax}

\begin{table*}[h]
\centering 
\begin{tabular}{c c c l} 
(program) & (P)  & ::= & $\textnormal{state}_1, \textnormal{state}_2, ..., \textnormal{state}_n $ main \\ 
(state definition) & (state) & ::= & state S case of S \{ $\overline{\textnormal{d}} \}$\\
(declaration) & (d) & ::=  & $\textnormal{method} \mid \textnormal{field} \mid \textnormal{state} \mid \textnormal{pts-def} \mid \textnormal{pts-inst}$  \\
(method-decl) & (method) & ::= & \colorbox{gray!50}{$\tau_r$ $m_i$ ($\overline{\tau_{ai} \gg \tau_{ai'} a_i}$)[$\overline{\tau_j \gg \tau_{j'} a _j}$] \{{ field; method; stmt; e }\}} \\
(field-decl) & (field) & ::= & (var $\mid$ val) $\tau$ f \\ 
(p-typestate-def) & (pts-def) & ::= & \colorbox{gray!50}{type $\gamma \Pi_{(\phi: \Psi_{\mathpzc{P}}, s : state)}.\tau$} \\
(p-typestate-inst) & (pts-inst) & ::= & \colorbox{gray!50}{$\gamma$ ($\phi_i, s_i$)} \\
(statement) & (stmt) & ::= & let x = e in stmt \\
		& & &	    $\mid$ let \^{x}.f = e in stmt \\
	& (state-change) & &	    $\mid$ e $\leftarrow$ e in stmt \\
	& (match) & &	    $\mid$ match (e) $\overline{\textnormal{case e \{e\}}}$ \\
 		& & &	    $\mid$\colorbox{gray!50}{while [$\exists.\phi$] ($e_1 : \textnormal{bool}$, $e_2$)} \\
		& & &	    $\mid$ case e \{ e \} \\
		& & & 	    $\mid$ \colorbox{gray!50}{skip }\\
(expression) & (e) & ::= & x $\mid$ \^x $\mid$ new S() \colorbox{gray!50}{$\mid$ new S ($\phi : \Psi_{\mathpzc{P}}$)}\\
		& & &	    $\mid$ e.m($\textnormal{e}_1, \textnormal{e}_2,..., \textnormal{e}_p$) \\
		& & &	    $\mid$ e ; e \\
		& & &	    $\mid$ c \\
(const)     & (c) & ::= & boolliteral $\mid$  intliteral $\mid$ stringliteral \\
& & & \\
(variable name) & x , \^{x} this & & \\
(field name) & f & & \\
(method name) & m , main& &  \\
(type family name) & $\gamma$ & &  \\
(state name) & S & & \\ 
(abstract locations) & $l_i$ & & \\

\hline

\end{tabular}
\caption{Core Language Syntax, \colorbox{gray!50}{shaded} represents related to p-typestate } 
\label{table:syntax} 
\end{table*}


The core language syntax is defined in Table~\ref{table:syntax}. The language is a typestate-oriented language, similar to a simple object-oriented language core like Feather Weight Java~\cite{FWJ}, apart from states and other features included to make p-typestates as first class members. A valid program in the language is a set of state definitions followed by a \textsf{main} method. A state definition (\textbf{state}) models a regular typestate state. State definition has a set of declarations (represented by $\overline{\textnormal{d}}$). A declaration (\textbf{d}) can be any of the method, field, state, p-typestate property definition or a p-typestate instantiation. Consider Figure~\ref{fig:brtypestatexmlparser} again showing a p-typestate program for \textsf{XMLParserSimple}. We will be referring to this example while explaining the syntactic features of the language. 

For brevity, we only present the features which differ from a general object based calculus~\cite{FWJ}. Method declarations define p-typestate transition relations over the operation of the language. A method declaration (\textbf{method}) contains \textit{method type} defining method structure and \textit{method body}. A \textit{method type} includes method return type $\tau_r$, method name $m_i$, list of typed formal parameters shown using $\overline{\tau_{ai} \gg \tau_{ai'}} a_i$ and a list of typed environment fields, represented by $\overline{\tau_j \gg \tau_{j'}}a _j$. The first argument in the environment list is by default, the base object of the method call. Each declaration of the form $\tau_i \gg \tau_{i'} a$ represents the p-typestate state change contract for $a$. It can be viewed as $Pre_{op, i}$ and $Post_{op, i, k}$ sets defined earlier in the formal definition of p-typestate for the type of variable $a$. For instance, 
consider method definition for \textsf{startReading} defined in \textsf{XMLParserSimple} state in Figure~\ref{fig:brtypestatexmlparser}(line 6). The p-typestate annotation for base object of the method is defined as, [\textsf{unique SafeSimpleParser (n, m) $\rightarrow$ Close $\gg$ unique SafeSimpleParser (n, m) $\rightarrow$ Open}]. The annotation represents pre- and post- p-typestate condition for the \textsf{startReading} method. It enforces that the base object is in \textsf{Close} (ClosedParser) state with integers n, m (n and m represent the number $\langle el \rangle$ and $\langle /el \rangle$ scanned respectively by the parser) for the call to be valid. Post method execution, the typestate must be toggled to \textsf{Open} and values for integers n, m must remain unchanged.

A \textit{method body} is a sequential composition of fields, methods, stmt declarations and expressions. A p-typestate-def (\textbf{pts-def}) is a declaration of a dependent type family indexed over Presburger formulas $\Psi_{\mathpzc{P}}$ and states of regular typestate and has a type name $\gamma$. It defines a new p-typestate property. For instance, line 3 in Figure~\ref{fig:brtypestatexmlparser} defines a p-typestate with name \textsf{SafeSimpleParser}, dependent on a Presburger formula over two integer variables \textit{ns, ne}, representing number of start and end $XML$ elements scanned respectively, and a state \textsf{Parser} $\in$ \{ \textsf{Open, Close} \}. A p-typestate-inst declaration (\textbf{pts-inst}) is an instance of a p-typestate property $\gamma$ with a Presburger formula $\phi_i$ and a state $s_i$. Intuitively, each such instance of a p-typestate-def captures a state of the p-typestate property satisfying property $\phi_i$ and having a regular typestate $s_i$. For example, Figure~\ref{fig:brtypestatexmlparser} shows a p-typestate-inst at line 24 for the p-typestate-def \textsf{SafeSimpleParser} defined earlier at line 3. Statements (\textbf{stmt}) in the core language are usual statements with a few special ones. (\textbf{state-change}) lets us define explicit p-typestate changes to an object. 

Another interesting statement is the (\textbf{while}) statement, which is like a usual while in other languages with one key addition. The while statement assumes an annotated loop invariant Presburger formula $\exists.\phi$. This assumption is essential for termination guarantee of our type-checking algorithm. The invariant can be provided by the programmer or can be inferred/calculated automatically. In Section~\ref{inference} we present a novel technique for such a loop invariant calculation. A typical while statement in our language looks like line 36 in Figure~\ref{fig:brtypestatexmlparser}. The (\textbf{match}) and (\textbf{case}) statements are analogous to conditional \textit{switch case} statements in typical imperative languages.
The (\textbf{new}) expressions in the language are similar to those in other Object oriented languages with the addition of \textbf{new S($\phi : \Psi_{\mathpzc{P}}$)}, which initializes an object of a state S with initial constraints $\phi$. \\

\subsection{Types}
\label{subsec:types}
\begin{table*}[h]
\centering 
\begin{tabular}{c c c l} 
(type) & ($\tau$)& ::= &  void $\mid$ int $\mid$ bool $\mid$ string \\
(state)&&&				$\mid$ S \\
(p-typestate transition)&&&				$\mid$ \colorbox{gray!50}{$\tau_i \gg \tau_j$} \\
(function type)&&&				$\mid$ $\tau_1 \rightarrow \tau_2$ \\ 
(method type)&&&				$\mid$ \colorbox{gray!50}{$\tau_1 \rightarrow \tau_2$ [$\overline{\tau_i \gg \tau_j}$]} \\
(permission type)&&&				$\mid$ (a, $\tau$) \\
(p-typestate property) &&&				$\mid$ \colorbox{gray!50}{$\Pi$ ($\phi : \Psi_{\mathpzc{P}}$, s : S).$\tau$} \\
(p-typestate state )&&&				$\mid$ \colorbox{gray!50}{($\phi$, s).$\tau$} \\
(permission) & (a) & ::= & unique  $\mid$ immutable \\
(type context) & ($\Gamma$) & ::=  & $\bullet$ $\mid$ $\delta$, $\Gamma$	\\
(type map)& ($\delta$) & ::=  & x : $\tau$ $\mid$  e : $\tau$ $\mid$ d : $\tau$ $\mid$ P : $\tau$ $\mid$ \colorbox{gray!50}{$\tau$ : $\star$}  \\
(heap) & ($\Theta$) & ::=  & $\bullet$ $\mid$ $\theta$, $\Theta$	\\
 & ($\theta$) & ::=  & x, \^x $\mapsto$ value  \\
(value) & value & ::= & $\rho$ $\mid$ c $\mid$ d $\mid$ new S() $\mid$ new S ($\phi: \Psi_{\mathpzc{P}}$) $\mid$ $l_i$ \\
(p-typestate-parameter) & $\Psi_{\mathpzc{P}}$ & ::= & $\phi$ \\
 & & & \\
(Presburger Formula) & $\phi$ & ::= & b $\mid$ $\phi_1 \wedge \phi_2$ $\mid$ $\phi_1 \vee \phi_2$ $\mid$ $\sim \phi$ $\mid$ $\exists v. \phi$ \\

(Boolean Expression & (b) & ::= & true $\mid$ false $\mid$ i == j $\mid$ $i \leq j$ $\mid$ $i \geq j$ $\mid$ $i \neq  j$ $\mid$ i == int \\

(Arithmetic Expression) & (i) & ::= & c $\mid$ v $\mid$ c * a $\mid$ $i_1$ + $i_2$ $\mid$ - i \\
(Index Context) & ($\Phi$) & ::=  & $\bullet$ $\mid$ $\phi$, $\Phi$	\\
\hline
\end{tabular}
\caption{Core Language Types and Context} 
\label{table:type} 
\end{table*} 
The core-language has a rich set of types with normal as well as dependent types which represent p-typestate property and its instances.
The type syntax, type context and index environment are presented in Table~\ref{table:type}. A type ($\tau$) can be a constant primitive type \textit{viz.}, void, int, bool or string type. Each state S is also a valid type. A (\textbf{p-typestate property}) is a dependent function type family representing a parameterized typestate and (\textbf{p-typestate state}) is an instance of this type family represented by $(\phi, s).\tau$, representing a state of the p-typestate property, satisfying formula $\phi$ and having a regular typestate state $s$. 
There are special composite types like, (\textbf{p-typestate transition}) $\tau_i \gg \tau_j$, representing Hoare style Pre and Post p-typestates for an expression or declaration. This allows us to capture p-typestate change from $\tau_i$ to $\tau_j$. A (\textbf{function type}) $\tau_1 \rightarrow \tau_2$ is a normal function type from $\tau_1$ to $\tau_2$. A (\textbf{method type}) as presented in method declaration is a composite type, composed of function type and a list of p-typestate transition types for each possible transition of parameters and environment references. A (\textbf{permission type}) (a, $\tau$), represents a type annotated with a permission. A permission manages the references to an object, it can be either \textbf{unique}, representing a single unique reference to an object or \textbf{immutable}, representing at least one alias to the reference thus making the typestate of the object reference immutatble. The permission system is explained further in Section ~\ref{subsec:typingrules}.

A typing environment $\Gamma$ is either an empty environment or a list of type maps ($\delta$). These maps are pairs mapping each variable, expression, declaration, program or a type, to a type. The dynamic state of the language is captured by the heap $\Theta$, which is either an empty heap or a list of value maps ($\theta$), mapping value and reference variables (x, \^{x}) to values. A value can be either a constant, a declaration, an abstract location or an object defined by new expressions, or an abstract value $\rho$.

A p-typestate parameter set $\Psi_{\mathpzc{P}}$, defines the possible index set for p-typestate properties and its instances. It defines a family of Presburger arithmetic logical formulas. A Presburger formula has a standard definition of \textit{First Order} logical constraints with arithmetic addition over auxiliary integer variables. 
Finally the typing rules also maintain an Index Context $\Phi$, keeping track of Presburger definable constraints generated via various typing rules. The Index Context is either empty or a list of Presburger formulas.

\subsection{Operational Semantics of the Core Language}
We present a big step operational semantics for the core-language in the Appendix in view of the limited space. An operational semantics is indispensable for ensuring the correctness of the system designed using our core  language. An abstract state of the program is defined as a pair ($\Theta, \Delta$), two variable to value maps mapping reference variables to abstract locations and value variables to values respectively. The big step semantics are presented as judgments $(\Theta, \Delta) \vdash e : \rho; (\Theta', \Delta')$. Such a judgment states that an expression $e$ evaluates in the program state $(\Theta, \Delta)$, to an abstract value $\rho$ and changes the program state to $(\Theta', \Delta')$ in the process. If the expression does not evaluate to a value (like, statements), the judgment drops the returned value $\rho$. Interested readers should refer Appendix, Section~\ref{semantics-long} for these semantic rules in Figure~\ref{fig:opsemantics1} and~\ref{fig:opsemantics2} along with their detailed explanation.

\subsection{Type System and Typing Rules}
\label{subsec:typingrules}
The p-typestate type system and corresponding typing rules enforces a p-typestate property, and aides in writing correct by construction programs which enforce protocol \textit{implementation} and \textit{usage} correctness. In practice, the typing rules inductively generate and transform a set of constraints as Presburger formulas, capturing some p-typestate property. These constraints are then passed to a simple SMT solver like Z3~\cite{z3}, which then checks for the validity of the constraints to verify the correctness of the program or returns a counter example showing a typestate violation.

\begin{figure*}[h]
\begin{center}
\hspace*{5ex} \inference[T-Pts-F]{ \Phi , \Gamma \vdash t : \star & \Phi, \Gamma \vdash {\phi : t} & \Phi, \Gamma;(\phi : t) \vdash S : \star \\ t \in \Psi } { \Gamma , \Phi \vdash \Pi({\phi : t, s : S}).\tau : \star } 

\bigskip

\end{center}

\begin{center}
\hspace*{5ex} \inference[T-Pts-I]{ (\Phi, \Gamma) \vdash {\phi : t} & (\Phi, {\phi : t}), \Gamma \vdash s : S    } { (\Phi, \Gamma ) \vdash \textnormal{type $\gamma$} \Pi(\phi : t, s : S).\tau : \Pi({\phi : t, s : S}).\tau } 

\bigskip

\end{center}

\begin{center}
\hspace*{5ex} \inference[T-Pts-C]{ ( \Phi, \Gamma) \vdash \textnormal{type $\gamma$} (\phi : t, s : S).\tau : \Pi({\phi : t, s : S}).\tau \\  (\Phi, \Gamma \vdash \phi_c : t) & (\Phi, (\Gamma; {\phi_c : t})) \vdash s_c : S } {  (\Phi, \Gamma) \vdash [\phi_c / \phi, s_c / s]\tau : (\phi_c, s_c).\tau  } 

\bigskip

\end{center}  

\begin{center}
\hspace*{5ex} \inference[T-mcall]{ (\Phi, \Gamma) \vdash e : (\Phi_1, \tau_b) & \tau_b =  (\phi_b, S_b).\tau \\
 mtype(m , S_b) = \textnormal{$T_r$ m($\overline{T_{i} >> T_{i}'} a _i$)[$\overline{T_{this} >> T_{this}'}$]}\{ e_m \} \\
(\Phi_1, (\Gamma; e : \tau_b) \vdash \tau_b <: T_{this} & (\Phi_1, (\Gamma; e : \tau_b) \vdash \overline{e_i: (\Phi_i, \tau_i)}  
\\ ((\Phi_1 \wedge (\bigwedge_{i} \Phi_i) (\Gamma; e : \tau_b; \overline{e_i : \tau_i)}) \vdash e_m : (\Phi_r, T_r) & \overline{\tau_i <: T_i}} 
{(\Phi,\Gamma) \vdash \textnormal{e.m($e_1, e_2, ... e_p$)} : (\Phi_r, T_r)}

\bigskip

\end{center}  
\begin{center}
\hspace*{5ex} \inference[T-Pts-update]{ (\Phi, \Gamma \vdash e_1 : (\Phi_1, \tau_1) \\ ( \Phi_1, (\Gamma, e_1 : \tau_1)) \vdash e : ( \Phi_2, \tau_1) } { (\Phi , \Gamma) \vdash \textnormal{e $\leftarrow$ $e_1$} : (\Phi_2, \tau_1)}

\bigskip 
\end{center}  
\caption{Selective Typing Rules}
\label{fig:selective-typing}
\end{figure*}

The full p-typestate typing rules are presented in Appendix in view of the limited space. Here we present a few interesting rules from the perspective of parameterized typestates, refer Figure~\ref{fig:selective-typing}. Each typing judgment in these rules is of the form \textit{$(\Phi, \Gamma) \vdash e : (\Phi', \tau)$}. It states that in the given typing context $\Gamma$ and dependent index terms constraint environment $\Phi$ (ref. table~\ref{table:type}), the expression $e$ is well typed and has a type $\tau$ and the typing of the expression updates the $\Phi$ to $\Phi'$. Any well formed type has a kind which we model as $\star$ in our type system. The p-typestate formation rule (T-Pts-F) defines a p-typestate property parameterized by a Presburger formula $\phi$ and a regular typestate state $s$. The p-typestate introduction rule (T-Pts-I) allows to introduce a concrete p-typestate property as a dependent type family. For instance, Figure~\ref{fig:dep-family-typing} (line 9) shows an introduction of a p-typestate property as a dependent type family named \textsf{depState} indexed with an invariant over two counter variables $\textsf{m, n}$, representing number of items produced and consumed respectively from a channel, and a state of the channel, \textsf{Channel} $\in$ $\{ Open, Close \}$. The p-typestate computation rule (T-Pts-C) defines an instantiation of a p-typestate property by an appropriate Presburger formula and a concrete regular typestate state. For example, Figure~\ref{fig:dep-family-typing} (line 13) creates a term named \textsf{safeChannel}  of the type ($ m=0, n=0, Open $).\textsf{depState}. This represents a state ($\forall m,n. m=0, n=0, m \geq n, Open$) of the p-property \textsf{depState}. Besides these, other interesting rules are related to p-typestate transitions, like (T-mcall) and (T-update). The (T-mcall) rule checks the validity of a method call, it ensures that p-typestate associated with each actual parameter and environment variable satisfies the Pre- p-typestate annotation for the callee. It checks the return type and updates the p-typestates for each parameter and environment variables according to the post- p-typestate annotations. The (T-update) rule allows an explicit p-typestate change for an expression. Please refer Appendix (Section~\ref{subsec:type-system}) for a complete list of typing rules.

\begin{figure}
 \begin{lstlisting}[xleftmargin=\fboxsep, xrightmargin=-\fboxsep, basicstyle=\selectfont\ttfamily\scriptsize, numbers=left, extendedchars=true, numberstyle=\tiny, mathescape=true ] 
(...)
state Channel{
	var Integer produce;
	var Integer consume;
	(...)
}
state DependentNew{
	//($\Phi$ = $\bot$ , $\Gamma \vdash e$ )/* Presburger contsraints $\Phi$ and Typing environment $\Gamma$ */
	type depState : Pi (m , n | m >= n ) -> Channel;
	 // ($\Phi'$ = m >= 0 $\wedge$ n >=0 $\wedge$ m >= n, $\star$) 
	method void main(){
	  // ($\Phi$ = $\forall$m, n. m >= 0 $\wedge$ n >=0 $\wedge$ m >= n, $\star$) 
	  var unique depState (m=0, n=0) -> Open safeChannel = new Channel{produce = 0; consume=0};
	  // ($\Phi'$ = $\forall$m, n. m = 0 $\wedge$ n =0 $\wedge$ m >= n, $depState(0,0, Open)$) 
	}
    }
}
\end{lstlisting}
\caption{Example code fragment, p-typestate introduction and instantiation}
\label{fig:dep-family-typing}
\end{figure}


\subsection{Handling Aliases}
Aliasing is a critical problem in handling typestates in object based languages. A sound typestate type checker must be aware of all the references to an object in order to capture any possible typestate transition. 
To track p-typestate changes in the presence of aliasing we use an \textit{access permission} system~\cite{Bierhoff:2009}. An access permission records whether a given reference to an object is \textit{unique}, in which case it may be used to change the typestate of the system, else if the object is shared amongst more than one references, we assign it an \textit{immutable} permission and do not allow typestate changes to the object's typestate via this reference. A \textit{unique} permission associated with a parameter guarantees to the method that it can access the object only through the current reference. It guarantees to the caller of the method that upon return, the method would not have created an alias for the parameter object. An operation can downgrade the \textit{unique} permission of a reference to \textit{immutable}. The reader should note here that the permission system is simple and can be refined based on other works~\cite{adoption-and-focus}, we leave a more advanced permission system capable of capturing more precise p-typestate transitions as a part of future work.

Besides access permissions, the aliasing problem can also be handled by introducing linear types~\cite{Lafont,Wakeling} in the language. Linear type systems~\cite{Wadler90lineartypes,once-upon-a-type} give programmers explicit control over memory resources. Most basically, linear types require a critical invariant that every linear value is used exactly once. This makes it possible to use  linear types to manage and reason about aliases~\cite{linear-types-for-aliased-resources}. Linear types extension can be a suitable replacement for the current permission based type system. We leave both these extensions of our type system to handle aliasing as future work. 

\subsection{Soundness}
Although we present no formal proof of soundness, we believe, the p-typestate system to be sound. We leave the proof of soundness for future work. The soundness is defined as follows-
\paragraph{\textbf{Soundness}:}
Given a program with an input p-typestate property automaton $PTS$ encoded as method contracts, and possibly a \textsf{main} method defined as an interprocedural control flow graph with statements and expressions as nodes. A type correct program guarantees that each method implementation satisfies its contract (\textit{protocol implementation verification}) and the main method is p-typestate verified (refer definition ~\ref{def:pts-verification}) for the encoded property $PTS$.

\section{Calculating Loop Invariants}
\label{inference}
One of the most important challenges in automatic, inductive type checking of expressive type systems like Liquid types~\cite{liquid}, other refinement types, and our p-typestate type system is the requirement to annotate loops and recursive data structures with invariants. These invariants allow to generate modular proofs of correctness for the program in Floyd-Hoare style proofs of program correctness, and are fundamental for the termination guarantee of the p-typestate type checking algorithm. Till now we have assumed that loop invariants are being provided by the programmer, unfortunately, its a daunting task even for experienced programmers to provide such invariants. Moreover, in many cases the loop invariants provided by the programmer may be too weak and insufficiently precise to verify a given p-typestate property. Thus, it is important to be able to automatically calculate adequate inductive loop invariants for programs whenever possible. To tackle this challenge we present a novel and simple loop invariant calculation approach based on loop acceleration technique for Presburger definable transition systems~\cite{Flat-acceleration-symbolic,FAST}. 

\subsection{Calculating loop-invariants using acceleration for Presburger-Definable Transition Systems}

\subsubsection{Acceleration:}
The problem of calculating the reachable set of states (REACH) for an infinite state system is undecidable in general. Thus, model checking infinite state systems requires ``symbolic'' approach. This involves abstracting a symbolic model of the model checking problem and manipulating it to calculate fixpoints for forward and backward reachability sets. 
A naive fixpoint calculation for these infinite systems may diverge in general and thus has low probability of termination. 
Acceleration~\cite{Flat-acceleration-symbolic} is a popular technique which makes the convergence of the fixpoint calculation for such systems more frequent. The technique is analogous to abstract widening operation from the abstract interpretation domain.

\begin{definition}[Acceleration over a path $\pi$]
\textnormal{
Given a transition system $\mathbb{T}$ = $\langle Q, \Sigma, \Psi, \delta \rangle$ and a sequence of action $\pi \in \Sigma^{*}$. \textit{Acceleration} of $\pi$ over $\mathbb{T}$ is called $\pi^{*}$ -acceleration and is defined as a relation $\textit{Acc}_{\pi}$ $\subseteq (Q \times Q)$ such that (s, s') $\in$ $\textit{Acc}_{\pi}$ iff $\exists k \in \mathbb{N}.$ such that s $\xrightarrow{\pi^{k}}$ s'. We say that s' $\in$ $post_{\mathbb{T}}(\pi^{*}, s)$ or simply $post^{*}(s)$. The definition could be extended to a set of starting states S, by calculating $post^{*}(s), \forall s \in S$. The acceleration relation $\textit{Acc}_{\pi}$ is called $\pi$ acceleration or just acceleration when the context is obvious.}
\end{definition}

\subsubsection{Computing \textnormal{REACH} using acceleration:}
The acceleration set defined above can be effectively utilized to calculated REACH for a system. For a given subset of initial states or configurations X of the system, and a language $\mathbb{L} \subseteq \Sigma^{*}$, we define $post(\mathbb{L}, X) = \{ x' \mid \exists x \in X \wedge (x, x') \in Acc_{\pi} \wedge \exists \pi \in \mathbb{L}\}$. The set $post(\Sigma^{*}, X)$ of all states/configurations reachable from X(initial set of configurations) is defined as the reachability set REACH of the system.

\subsubsection{Using \textnormal{REACH} for loop invariant calculation:}
We model the loop invariant calculation for a program enforcing a p-typestate property as REACH finding problem over the counter system induced by the looping construct in the program. This reduction allows us to use known acceleration based reachable states computation approaches and tools. We use a Flat acceleration tool FAST~\cite{FAST} in our implementation to calculate REACH for \textsf{while} loops in programs. The reduction is straight forward, Presburger formulas over integer variables of the counter system for the input loop forms the symbolic domain and restrictions on the structure of these counter system guarantees the termination of the FAST tool. FAST calculates a Presburger definable formula representing the REACH set for the input loop with the given initial set of states. We use this formula as a loop invariant in p-typestate type checking to generate a modular proof of correctness of the program. 

The approach assumes that the input counter system for the loop is \textit{finite linear}~\cite{FAST} and \textit{flattable}~\cite{Flat-acceleration-symbolic}. These restrictions are fundamental to the termination of the approach which uses a semi-algorithm to calculate REACH set. The tool's acceleration algorithm can run on any finite linear system and is a complete procedure for flattable finite linear systems. It provides no termination guarantee for other general class of finite linear but non-flattable systems. 

\begin{definition}[Integer Counter System for Loops]
A counter system $C$ for a loop is a transition system, defined as a tuple $\langle Q, \Sigma, \Psi_{\mathpzc{P}}, \delta \rangle$. Where $Q$ is a finite set of states, $\Sigma$ is a finite set of Presburger definable actions. These actions simulate the p-typestate contracts associated with statements and expressions of the program. $\Psi_{\mathpzc{P}}$ is a Presburger definable set over $m$  integer variables $\mathbb{V}^{m}$ used in the loop.  $\Psi_{\mathpzc{P}}$ defines \textit{guards} for transitions given by $\delta$ : $(Q \times \Sigma \times \Psi_{\mathpzc{P}}) \mapsto (Q \times \Psi_{\mathpzc{P}})$. 
\end{definition}

A \textit{state} of $C$ is defined as a tuple $\mathbb{Z}^{m}$ assigning values to $\mathbb{V}^{m}$. A state $s_i$ satisfies a Presburger guard $\phi \in \Psi_{P}$ iff  $s_i \vDash \phi$.

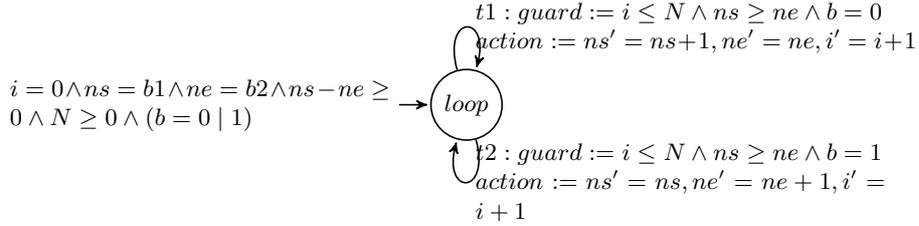
\begin{figure*}[htbp]
\centering
\begin{tikzpicture}[->,>=stealth',shorten >=1pt,auto,node distance=3cm,
                    semithick]
  \tikzstyle{every state}=[fill=white,draw=black,text=black]

  \node[initial,state, initial text={\parbox{5cm}{$i = 0 \wedge ns = b1 \wedge ne = b2 \wedge  ns - ne \geq 0 \wedge N \geq 0 \wedge (b = 0 \mid 1) $}}] (A)                    {$loop$};

  \path  (A) edge [loop above, text width=5.8cm, right] node {$ t1 :guard := i \leq N \wedge ns \geq ne \wedge b = 0 \newline action := ns' = ns + 1, ne' =ne, i' = i + 1$} (A);
    
  \path  (A) edge [loop below, text width=5.5cm, right] node {$ t2 :guard := i \leq N \wedge ns \geq ne \wedge b = 1 \newline action := ns' = ns, ne' =ne + 1, i' = i + 1$} (A);

\end{tikzpicture}
 \caption{Loop counter system for the while loop of Figure~\ref{fig:brtypestatexmlparser}}
\label{loopautomata}
\end{figure*}

Figure~\ref{loopautomata} shows the loop counter system for the \textsf{while} loop in the code fragment for \textsf{XMLParserSimple} defined in Figure~\ref{fig:brtypestatexmlparser}. 

\begin{figure*}
\centering
\includegraphics[scale=0.30]{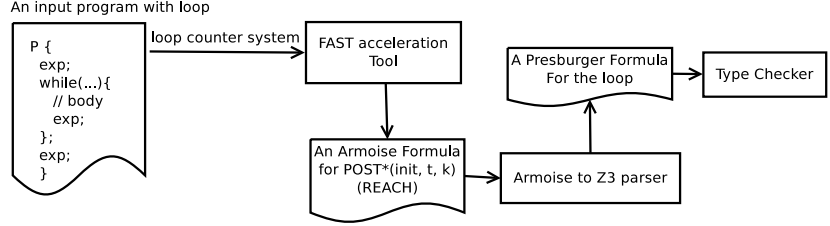} 
\caption{Block diagram for loop invariant calculation approach}
\label{flowinfer}
\end{figure*}

\subsubsection{Complete Approach:}

Figure~\ref{flowinfer}, shows a block diagram for the invariant calculation approach. The approach requires an integer counter system for the input loop. The counter system can be extracted by analyzing the loop body for pre- and post- p-typestate annotations of expressions and statements and their semantics. In this work we manually construct these counter systems for input loops. 

This loop counter system is passed as input to the FAST loop acceleration tool which generates (on termination) the REACH set for the loop as a Presburger formula in the representation of FAST output language Armoise~\cite{armoise}. We built an Armoise to Z3 parser which generates corresponding Z3 formula, and passes it to the type-checker. The type-checker composes the formula with the incoming constraints using the typing rule for \textsf{while} expression. We finally discharge the composed collective constraints to the Z3 solver. 

\paragraph{Example:}
Consider Figure~\ref{fig:brtypestatexmlparser} from the Overview section, 
Figure~\ref{loopautomata} shows the loop counter system for the \textsf{while} loop in the code fragment. The counter system defines the initial set of states, where $i$ is the counter variable for the loop (corresponding to the var \textsf{numberscanned} in the program), $b$ is a boolean variable representing the conditional variable in the loop body. The counter system also defines transitions $t1$ and $t2$, which are Presburger definable actions over index variables. The simplified loop invariant calculated by the approach in this example is ($\forall$ (\textsf{ns, ne}, $b1, b2, N) \in (nat, nat, nat, nat, nat)$. \textsf{ns} $\leq b1 + N \wedge$ \textsf{ne} $\leq b2 + N \wedge$ \textsf{ns - ne} $\geq$ 0 $\wedge$ \textsf{ns} $\geq b1 \wedge$ \textsf{ne} $\geq b2$). This loop invariant is indeed adequate to generate an inductive proof of correctness for the \textsf{main} method and prove the \textsf{main} method implementations in Figure~\ref{fig:main}(a) as correct and ~\ref{fig:main}(b) as incorrect respectively.

\begin{figure*}[htbp]
\begin{minipage}{0.5\textwidth}
\begin{lstlisting}[xleftmargin=\fboxsep, xrightmargin=-\fboxsep, basicstyle=\selectfont\ttfamily\scriptsize, numbers=left, extendedchars=true, numberstyle=\tiny, breaklines=true, mathescape=true]
state Broadcast{
  method sendBroadcast (_ >> _)[_ >> _]{
    var x, y, z, N;
    x = N; y = 0; z = 0;
    while(true){
	match (*){
	case (x >= 1) { x = x + y -1; y = z + 1; z =0; } 
	  case (y >= 0) { y = y -1; z = z + 1; }
	  }default {}	
	 };
    };
  }
}
\end{lstlisting}
\caption{\textbf{Example: Simple Broadcast} Approach fails to calculate REACH}
\label{non-working-inferloop}
\end{minipage}
\hspace{0.2cm}
\begin{minipage}{0.5\textwidth}
\centering
\begin{tikzpicture}[->,>=stealth',shorten >=1pt,auto,node distance=3cm,
                    semithick]
  \tikzset{
    state/.style={
           rectangle,
           draw=black, thick,
           minimum height=2em,
           inner sep=2pt
           },
}
   
  \node[state] (QUERY) 
  {\begin{tabular}{l}
 \parbox{4cm}{
    main()\{ \\
	\hspace*{5mm}$\phi_{pre}$ \\
 	\hspace*{5mm}$\textnormal{S}_1$; \\
	\hspace*{5mm}$\phi_{in}$ \\
	\hspace*{7mm}$\longleftarrow \phi$ \\
	\hspace*{5mm}while (b) \{ \\
	\hspace*{7mm}$\textnormal{S}_2$; \\
	\hspace*{5mm}\} \\
 	\hspace*{5mm}$\textnormal{S}_3$; \\
	\hspace*{5mm}$\phi_{post}$\\
	\}
    }\\[4em]
  \end{tabular}};

\end{tikzpicture}
 \caption{Loop invariant $\phi$ (specified or calculated) for a program with a \textsf{while} loop }
\label{fig:inductive-invariant}
\end{minipage}
\end{figure*}

It is worth mentioning that loop invariant calculation approach might fail to terminate and return the Presburger definable REACH set for several reasons: (1) The loop counter system is not a \textit{finite linear system}. (2) The counter system is not \textit{flattable}. (3) The approach timed out (for time limit $>$ 3 min). Reasons (1) is ruled out by the assumption of the input loop counter system. The major reason for non-termination therefore is either (2), which although is assumed, but cannot be verified as checking \textit{flattability} of a Counter system is undecidable~\cite{FAST}, or, in certain cases the termination is not reached in practical time limits due to (3). Figure~\ref{non-working-inferloop} presents an example(adapted from ~\cite{FAST}) code fragment in our core language with a while loop and two inner conditional expressions for which the approach fails to terminate due to reason (2). More results showing both successful calculations and failures due to reasons (2) or (3) are discussed in section~\ref{results}.

\subsubsection{Completeness of the loop invariant calculation approach}
Given a p-typestate annotated program $P$ with a \textsf{while} loop, we present a claim for the ``completeness'' of our approach.

Consider Figure~\ref{fig:inductive-invariant}, Each $\textnormal{S}_{i}$ is a block of statements or expressions causing p-typestate constraints changes as depicted in the program. Thus $\textnormal{S}_1$ when executed in a pre- ptypestate with associated constraint $\phi_{pre}$, updates the constraints to $\phi_{in}$. We abuse the notations here and simply write this transition as $(\phi_{pre} ; S_1) \Rightarrow \phi_{in}$. The formula $\phi$ represents the loop invariant (specified or calculated) for the \textsf{while} loop.

\begin{definition}[Adequate Inductive Invariant]
\textnormal{We say the invariant $\phi$ is an \textit{adequate inductive invariant} for the given program specification, if following conditions hold-
\begin{itemize}
 \item $\phi_{in} \Rightarrow \phi$ (``inductive invariant'')
\item $((\phi \wedge b ) ; S_2) \Rightarrow \phi$ (``inductive invariant'')
\item $((\phi \wedge \neg b ) ; S_3) \Rightarrow \phi_{post}$ (``adequate'')
\end{itemize}}
\end{definition}

\subsubsection{``Completeness'' claim:}
If our computation of a loop invariant via FAST succeeds, we are guaranteed that it is the ``best'' possible invariant: if the proof/type-checking does not succeed with this invariant, it cannot succeed with \emph{any} other invariant. This is essentially because this particular loop acceleration technique returns (if it does terminate) the \emph{exact} set of reachable states at the loop head. To substantiate our claim, consider the program structure in Figure~\ref{fig:inductive-invariant}, and suppose FAST returns an invariant $\phi$. Suppose further that it is not sufficient (i.e.\@ one of the three conditions above fail). Now by virtue of the fact that $\phi$ represents the exact set of reachable states at the loop head, it is easy to see that it will satisfy the inductiveness conditions. So it must be the adequacy condition which fails. That is, there exists a state $s$ satisfying $\phi$ and $\neg b$, such that after doing $S_3$ we reach a state that does not satisfy $\phi_{post}$. But any other proposed
invariant $\psi$ must include $\phi$, if it satisfies the inductiveness conditions. Hence $s$ must be included in $\psi$ as well, and $\psi$ will similarly fail the adequacy check.


\section{Results}
\label{results}

In this section we present a small set of useful programs with important non-regular program properties which cannot be modeled using regular typestate but we have statically verified them using p-typestates. 
\begin{table*}[htbp]
	\centering 
	\begin{tabular}{ |p{2.5cm}| p{8cm}| P{1.5cm} |} 
		\hline
		 Property Name & \centering Property Description & LI*\\
		\hline
		 Producer-Consumer &Producer(P), Consumer(C), SharedChannel(sc) \newline 
				     items produced (np), items consumed (nc) \newline
				     $S$(sc) = \{open, close\}, sc.produce() iff open, sc.consume() requires iff \newline
				     $\phi$ = $\forall op \in {\{open, close, produce, consume\}}^{*}$. np $\geq$ nc & S \\
		  \hline
		 SizedArray and SizedList& Array sized N (A(N)), $i_{th}$ index access (A[i]) \newline
					  $\phi_{1}$ = $\forall A[i]$. $i \leq$ N \newline
					  $\phi_{2}$ = A.add() iff N $\leq$ $k \in \mathbb{N}$  & S\\
		  \hline
		
		 Binary Search& Binary search implementation suing SizedArray. No runtime array bounds checks needed  & S\\
		\hline
		List Reversal and Append & List Sized N1 (L1(N1)), List Sized N2 (L2(N2)) \newline 
					    L1 = [$x_1, x_2...,x_{N1}$], L2 = [$y_1, y_2...,y_{N2}$] \newline
					    L = [$x_1, x_2...,x_{N1}, y_{N2}, y_{N2-1}...,y_1$] \newline
					    $\phi$ = size(L) = N1 + N2 & S \\
		 \hline
		Banking Problem & Account (Acc), Amount Withdrawn (W), Amount Deposited (D) \newline
				$S$(Acc) = \{active, inactive\}, Acc.withdraw() iff active, Acc.deposit() iff active \newline
				$\phi$ = $\forall$ transactions. D $\geq$ W & S \\
		\hline
		 Train Running Protocol & Train speed control protocol ~\cite{Flat-acceleration-symbolic} & S \\
		 \hline 
		 StackModel \newline [Figure~\ref{fig:stackmodel}] & Simulating a Stack with pushm, pop, top, empty and acceptance using p-typestates & N/A \\
		 \hline
		
		 XMLParser & XMLParser (P), open tag $\langle el_i \rangle$ \newline  close tag $\langle /el_i \rangle$ \newline 
			    $\phi$ =$\forall \sigma \in {\{\langle el_i \rangle, \langle /el_i \rangle \}}^{*}$. $\sigma \in \textnormal{Dyck}_m$ & S\\ 
		  \hline
		  XMLParser- \newline Simple & XMLParser (P), open tag $\langle el \rangle$ \newline  close tag $\langle /el \rangle$ \newline 
			    $\phi$ =$\forall \sigma \in {\{\langle el \rangle, \langle /el \rangle\}}^{*}$. $\sigma \in \textnormal{Dyck}_2$ & S\\ 
		\hline
		 CFL-Parser \newline ($a^nb^n$ -Parser)& Given a graph G, with nodes \{ $n_1, n_2,..., n_N$ \} and each edge labeled $l \in \{ a, b \} $, $\pi(p)$ = string generated by labels of a path $p$ \newline
			 $\phi_1$ = $\forall p \in {\{a,b\}}^{*}$ s.t. $\exists$ a path $p$ b/w $n_i$ and $n_j$, s.t. $z$ = $\pi(p)$ $\Rightarrow$ \newline
							  $\phi_1$ = $z \in$ $a^nb^n$ \newline
							 $\phi_2$ = $\forall z'$, s.t. $z=z'x$, $z' \in a^jb^k$, where $j \geq k$ 
							  & S \\
		\hline
		 IVP checking & Given an interprocedural control G flow graph with edges labeled with method calls ($c_i$) and returns ($r_i$). \newline
			      $\phi_1$ =$\forall \sigma \in {\{el_i, /el_i\}}^{*}$. $\sigma \in \textnormal{Dyck}_m$ \newline
			      $\phi_2$ =$\forall \sigma'$ s.t. $\sigma = \sigma'.x$ $\textnormal{D}(\sigma') \geq 0$& S\\
		 \hline
		 Broadcast & Simple Broadcast example Figure~\ref{non-working-inferloop} & F \\
		\hline
		 Synchronized Inc/Dec & Integer counters increment and decrement between 2 threads. & TO ($>$ 3 mins.) \\
		 \hline
		
	\end{tabular}
	\caption{Statically Verified p-typestate Properties. *- LI = (Loop Invariant) S = Success, F = Failure, T/O = Timed Out, N/A = No loops} 
	\label{table:results} 
\end{table*} 

\subsection{Implementation}
The core-language parser has been implemented using \textit{JavaCC}~\cite{javacc} parser generator. The prototype p-typestate type checker has been implemented in Java and contains three major components. (i) A Presburger constraint generator which generates constraints based on p-typestate typing rules (ii) A loop invariant calculator, described earlier which uses FAST tool. (iii) A Presburger constraint solver.
The constraint generator has been written in Java. The loop invariant calculator requires an Armoise (internal formula representation language of FAST) to Z3 parser and this parser has been written in Java. We use Z3 solver for checking the validity the generated constraints. 

\subsection{Results}
Table~\ref{table:results} presents a set of statically verified p-typestate properties and invariants implemented in our language. The table gives property names, their short description (formal / informal) and reports the success, failure or time out for the loop invariant calculation approach for  the property, if applicable. Some of the property names are annotated with the figure representing its implementation. Each of these example properties took a few seconds for Presburger constraints generation by the type-checker and solving using Z3 on an Intel Xeon, 8 core CPU@2GHz with 3 GB of memory. The detailed definition of each of these properties and their implementation in our language may be found online at ~\cite{p-typestate}. We briefly explain a few of these here. The figures showing code snippets are moved to Appendix due to space constraints.
%
%

Figure~\ref{fig:sizedList} (please refer Appendix Section~\ref{subsec:result}) presents a sized version of a List with a p-typestate \textsf{SizedListTy}, capturing the size of a List. Each method of the \textsf{SizedList} state is annotated with a pre- and post- p-typestates.  For example, method \textsf{append} takes an input list of a p-typestate \textsf{(unique SizedListTy(m) $->$ List)}, representing a list of Size \textsf{m} and requires the size of base \textsf{List} object to be \textsf{n} and returns a \textsf{List} of size \textsf{m+n}. The type checking algorithm statically verifies these pre- and post p-typestate annotations. We have implemented sized versions of arrays and lists as a library where sizes are checked and enforced using the p-typestate type system.

Figure~\ref{fig:stackmodel} (please refer Appendix Section~\ref{subsec:result}) presents a bit intricate example. Here we simulate a stack using counters. The \textsf{StackModelType} represents a p-typestate, modeling a stack, with (\textsf{c1, c2}) representing two binary strings, such that, the least significant bits of \textsf{c1} and \textsf{c2} defines the top element of the stack. For example, if \textsf{c1} = ``0b00'' and \textsf{c2} = ``0b01'', then ``01'' is the stack top. Each stack element(\textsf{StackElementWithId}) has an Id, given by a binary string \textsf{b1b2}. Thus at any time, strings \textsf{c1, c2} define the contents of the stack. Pushing an element with Id (\textsf{b1b2}) is achieved by updating \textsf{c1} and \textsf{c2} to \textsf{c1'} and \textsf{c2'} respectively by multiplying each by 2 and adding \textsf{b1} and \textsf{b2} respectively. To pop an element, \textsf{c1, c2} are divided by 2. We use this StackModel to model and verify various important context-free properties like XMLParser, CFL-Parser, etc.

\section{Related Work}
\label{relatedwork}
Table~\ref{table:related} presents a comparison of our p-typestate and core-language against the other related works over different dimensions or features. The table clearly shows how p-typestate type system and core-language stand out over other related work.

\subsection{Language Support for Typestate}

Plaid~\cite{Plaid} includes regular typestates as first class objects in the language enabling it to define typestate properties directly in the language. It is useful but as a consequence of regular typestates, lacks expressive power to implement a non-regular properties. Modular typestate for Object-Oriented programs~\cite{typestates-for-objects} models regular typestates as predicates over object fields, and specially handles subclassing in objects. Authors of ~\cite{typestates-for-objects} briefly discuss expressive limitations of their typestates and allude to an extension of their typestate. Moreover, the work does not provide any concrete implementation of their type system and core language. Our work subsumes these typestate works as we can express regular typestate as a special degenerate case of p-typestate. Working out a modular p-typestate system following ~\cite{typestates-for-objects} can be interesting, we leave this as a possible future work. Extended Static Checking(ESC)~\cite{ESC} for Java provides an annotation language and a static checking of properties over Java programs. It is based on First Order Logic and general theorem proving which although gives expressiveness to the language but does not aim for decidable type-checking. Java Modeling Language (JML)~\cite{JML} lets the programmer describe typestate like properties as pre and post conditions of methods as specifications which are then compiled into runtime checks by the JML compiler. Although this allows it to model any non-regular program property, these properties can not be checked statically.

\begin{table*}[htbp]
	\centering 
	\begin{tabular}{ |p{5cm} | c | c | c | c | c |} 
		\hline
		 features $\downarrow$ $\diagup$ works $\rightarrow$ & p-typestate & X10~\cite{x-10} & DML~\cite{dml} & Xanadu~\cite{Xanadu} & $\textnormal{RT}^{*}$~\cite{typestates-for-objects,Plaid}\\
		 \hline
		{Typestates as first class members} & $\surd$& & & & $\surd$ \\
		Expressive Dependent Types & $\surd$ & $\surd$ & $\surd$ & $\surd$ & \\
		Decidable Type-checking &  $\surd$ &  & &$\surd$ & $\surd$ \\
		Efficient loop invariant calculation & $\surd$ & &  &  & \\
		Object oriented & $\surd$ & $\surd$  &   &   & $\surd$ \\
		\hline
		
	\end{tabular}
	\caption{Comparison with Related Works, *RT=Regular Typestate} 
	\label{table:related} 
\end{table*} 


\subsection{Fully Dependently Typed Languages}
 There are various fully dependently typed programming languages~\cite{Coq,Agda} which provide expressiveness to model rich non-regular program properties. This expressiveness comes at the cost of complexity of type-checking as type-checking is undecidable in general dependent type systems. A possible approach then can be to restrict the dependent terms of these languages to belong to a decidable theory. This is semantically equivalent to defining a language equivalent to ours (minus the p-typestate features) in these fully dependently typed languages like Coq or Agda. Further, to achieve this, a programmer needs to build a Presburger Theory in Coq, restrict the dependent terms to this theory and then write properties and programs in this restricted sub-languages. This encoding will be verbose and complex for a programmer to build and will further require a proof of decidability of type cheking. Moreover, it will still lack expressiveness to define rich p-typestate like pre- and post- conditions and p-typestate changes.

\subsection{Dependently typed extensions for languages}

Many works~\cite{DMLC,Xanadu,liquid,x-10} are inspired by dependent ML(DML)~\cite{dml} and provide some dependently typed extension of a non-dependently typed language. Our work differs from these in terms of being dependent extension of typestates, rather than of simple types (type associated with a data object remains constant throughout its life, while typestate is allowed to transit from pre- to post-condition), the language domain, decidability of type-checking, handling of loops and recursive data structures, etc. For example, DML adds dependent types to ML type system. It is a functional programming language, while our core language is an imperative, typestate-oriented dependently typed language. Our loop invariant calculation technique can be useful for type-checking general recursive functions in DML.

Xanadu~\cite{Xanadu} is an imperative version of the DML. Xanadu provides a static dependent type system and aims at eliminating array bound checks in imperative programs. Although Xanadu allows types to change during evaluation, it provides no support for specifying or verifying pre- and postconditions of methods, expressions and arguments which forms the basis of typesate like properties. Properties defined and verified by Xanadu can be easily specified and verified as p-typestate properties in our work. The loop invariant calculation approach presented by us is significantly different than their approach and obviates the need of loop invariant annotations from the programmer. Compared to this the ``Master type'' based ``state type'' calculation for loops in Xanadu requires annotation in certain cases or can be highly imprecise, otherwise.

The constrained type of X10~\cite{x-10} is related to the dependently typed language we provide in our work. X10 allows definition of ``constrained types'' which are dependent types with logical expressions over properties, final instance fields of a class, and final variables, in the scope of the type as dependent terms. It also allows different constraint systems as  compiler plugins. This again is a dependent extension of types rather than typestates.


An  X10 type may contain a final field or even a method call as an index of a type. This allows far more expressive constraints but requires symbolic execution to get the possible properties of these fields and method executions. Symbolic execution may lead to undecidable type checking.

X10 does not provide any automatic loop invariant calculation and does not describe how looping constructs are handled by the type checker. Finally, X10's type system allows conditional expressions to be a conjunction of different constraint families thereby making it more expressive, but does not discuss the decidability of the type checking over these complex expressions.

\section{Conclusion and Future Work}
\label{conclusion}

There are several directions for possible future work. Extension of p-typestates type system with gradual typing would aid in increasing the precision of the verification and will also aid in reducing the burden of annotation on the programmer. A modular p-typestate type system on the lines of Deline et. al. ~\cite{typestates-for-objects} will be really useful. Another interesting direction of future work will be to extend the p-typestate and core language with richer features like polymorphic types, concurrency, etc. This extension will lead to further investigation the connections between \textit{Session types}~\cite{session-types}, normal typestates and p-typestate. There are various non-regular properties over sessions which could benefit from such an extension. A complete formal characterization and applications of the loop invariant calculation approach presented in our paper will be another useful direction for future work.

\clearpage
\bibliographystyle{plain}
\bibliography{ref}

\section{Appendix}
\label{appendix}

\subsection{Operational Semantics of the Core Language}
\label{semantics-long}
 The abstract state of the program is defined as a pair ($\Theta, \Delta$), two variable to value maps mapping reference variables to abstract locations and value variables to values respectively. The big step semantics are presented as $(\Theta, \Delta) \vdash e : \rho; (\Theta', \Delta')$. Such a judgment states that an expression $e$ evaluates in the program state $(\Theta, \Delta)$, to an abstract value $\rho$ and changes the program state to $(\Theta', \Delta')$ in the process. If the expression does not evaluate to a value (statements), the judgment removes the returned value $\rho$. Figure~\ref{fig:opsemantics1} presents these semantic rules for the language. Some of these judgments are self explanatory, while the most interesting ones, most closely relevant to the typestate and p-Typestate are given by the rules \textbf{mcall, let, match, update, and while}. \textbf{mcall} has a call by value semantics. It checks that the receiver reference is mapped to a non-null (null is a special location) location and then creates an extended program state mapping each formal parameter expression $e_i$ to the values of the corresponding actual parameters and evaluates the body of the called method in this new extended state to change the state to $(\Theta_{out}, \Delta_{out})$. The \textbf{match} expression  evaluates the match expression $e$ and further evaluates each of the case expressions $e_i$ in this new program state returning $\rho_{ei}$, and possibly changing the state to ($\Theta_i, \Delta_i$). Since, the match expression could match to any of the possible case expression, we create an over-approximate value for state of the system post completion of the rule. Thus $(\Theta_{out}, \Delta_{out})$, is a union over all the state maps generated by each of the case expressions. The returned value $\oplus \rho_{ei}$ is one of the any possible returned value, thus this can bee seen as an indexed set of values, indexed over the case expression $e_i$. The \textbf{update} rule, refer Figure~\ref{fig:opsemantics2} evaluates the source expression $e'$ of the update expression, changing the state to $(\Theta', \Delta')$ and updates the fields of the target expression $e$, \{ $f_1, ... f_p$ \} by the values of the corresponding fields from the source expression $e'$. The final state is the new updated state with updated maps for each field of $e$ and the $e$ itself. The \textbf{while} rule semantics depend on the value of the conditional expression $b$, if the the condition evaluates to false (\textbf{while-false}) while updating the state to $\Theta', \Delta'$ during evaluation of the $b$, the expression evaluates the next expression (or statement) $e_n$ after the while body. The case for true condition (\textbf{while-true}) is much complex,which evaluates the body of the while statement $e$, in the updated environment and evaluates the next expression $e_n$, only in the new state ($\Theta', \Delta'$), which is obtained after a fix point for the loop is reached.

\begin{figure*}[htbp]

\begin{minipage}{.5\textwidth}
\begin{center}
\hspace*{5ex} \inference[const]{\Delta' = \Delta, (\rho = c) }{(\Theta; \Delta) \vdash c : \rho ; (\Theta, \Delta')}

\end{center}
\end{minipage}
\begin{minipage}{.5\textwidth}
\begin{center}
\hspace*{5ex} \inference[val-var]{\Delta' = \Delta, (x \mapsto \rho) & \rho = default(\Gamma(x)) }{(\Theta; \Delta) \vdash x : \rho ; (\Theta, \Delta')}

\end{center}
\end{minipage}

\bigskip

\begin{minipage}{.5\textwidth}

\begin{center}
\hspace*{5ex} \inference[ref-var]{\Theta' = \Theta, (\textnormal{\^x} \mapsto \rho) & \rho = default(\Gamma(\textnormal{\^x})) }{(\Theta; \Delta) \vdash \textnormal{\^x} : \rho ; (\Theta', \Delta)}

\end{center}
\end{minipage}
\hspace{0cm}
\begin{minipage}{.5\textwidth}
\begin{center}
\hspace*{5ex} \inference[de-ref]{ \Theta(\textnormal{\^x}) = \rho \\ \Theta(\rho) \neq null \\ \Theta(\rho) = new S (a_1 : \rho_1, ...,a_p : \rho_p) \\ \Theta' = \Theta[\textnormal{\^x}.f_j \mapsto \rho_j], (\textnormal{\^x} \mapsto \rho))) & \rho_j = \Theta(a_j) }{(\Theta; \Delta) \vdash \textnormal{\^x}.f_j : \rho_j ; (\Theta', \Delta)}

\end{center}

\end{minipage}
\bigskip

\begin{minipage}{.5\textwidth}
\begin{center}
\hspace*{5ex} \inference[new]{ \Theta' = \Theta, (\rho = new S() \mid new S (\phi))}{(\Theta; \Delta) \vdash new S() \mid new S (\phi) : \rho ; (\Theta', \Delta)}

\end{center}

\end{minipage}
\hspace{0cm}
\begin{minipage}{.5\textwidth}
\begin{center}
\hspace*{5ex} \inference[mcall]{\Theta(y) \neq null & \Theta(y) = \rho_m \\ \rho_m := \tau_r m (e_1, e_2,...e_p)[]\{ e_b \} \\ \Theta' = \Theta[e_i \mapsto \Theta(f_i)] \\
				   (\Theta' , \Delta \vdash e_b : \rho_b ;(\Theta'', \Delta'') )}{(\Theta, \Delta) \vdash y.m(f_1, f_2,...f_p) : \rho_b ; (\Theta'', \Delta'')}

\end{center}

\end{minipage}

\bigskip

\begin{center}
\hspace*{5ex} \inference[let]{ (\Theta, \Delta) \vdash e_1 : \rho_{e1}; (\Theta', \Delta') \\ \Theta'' = \Theta'[x \mapsto \rho_{e1}] \\ (\Theta'' , \Delta') \vdash stmt : \rho ; (\Theta_{out}, \Delta_{out})}{(\Theta; \Delta) \vdash \textnormal{let x = $e_1$ in stmt} : \rho; (\Theta_{out}, \Delta_{out})}

\end{center}

\bigskip

\begin{center}
\hspace*{5ex} \inference[match]{ (\Theta, \Delta) \vdash e : \rho_e; (\Theta', \Delta') \\ 
				    (\Theta', \Delta') \vdash e_i : \rho_{ei}; (\Theta_i, \Delta_i)\\
				    \Theta_{out} = \bigcup \Theta_i & \Delta_{out} = \bigcup \Delta_i \\
				    \rho_{out} \oplus \rho_{ei}}{(\Theta; \Delta) \vdash \textnormal{match e case $e_1 \{ b_1\}... e_p \{ b_p\}$} : \rho_{out} ; (\Theta_{out}, \Delta_{out})}

\end{center}

\bigskip

\caption{Big step operational semantics for the core language}
\label{fig:opsemantics1}
\end{figure*}

\begin{figure*}[htbp
]
 \begin{center}
\hspace*{5ex} \inference[update]{(\Theta, \Delta) \vdash e' : \rho_{e'}; (\Theta' , \Delta') \\
				    \rho_{e'} = new S_t ( f_1 : \rho_{t1}, f_2 : \rho_{t2}, ... f_p : \rho_{tp})\\
				      \Theta'(e) = \rho_e = new S_s ( f_1 : \rho_{s1}, f_2 : \rho_{s2}, ... f_p : \rho_{sp}) \\
					\Theta'' = \Theta [e \mapsto \rho_{e'}] \\
					  \Theta_{out} = \Theta'' [\forall \rho_e.f_i \mapsto \rho_{ti}]\\
					    \Delta_{out} = \Delta'}{(\Theta, \Delta \vdash e \leftarrow e' : \rho_{e'} ; (\Theta_{out}, \Delta_{out}))}

\end{center}

\bigskip
\begin{center}
\hspace*{5ex} \inference[while-false]{ (\Theta, \Delta) \vdash b : false; (\Theta', \Delta') \\ (\Theta', \Delta') \vdash e_n : \rho_n ; (\Theta'', \Delta'')}{(\Theta; \Delta) \vdash \textnormal{while [$\exists.\phi$] b \{ e \}} ; e_n : \rho_n ; (\Theta'', \Delta'')}

\end{center}

\bigskip

\begin{center}
\hspace*{5ex} \inference[while-true]{ (\Theta, \Delta) \vdash b : true; (\Theta', \Delta') \\ (\Theta', \Delta') \vdash e : \rho_e ; (\Theta'', \Delta'') \\ 
					(\Theta'', \Delta'') \vdash e_n : \rho_n ; (\Theta_{out}, \Delta_{out})}{(\Theta; \Delta) \vdash \textnormal{while [$\exists.phi$] b \{ e \}} ; e_n : \rho_n ; (\Theta_{out}, \Delta_{out})}

\end{center}

\bigskip

\begin{center}
\hspace*{5ex} \inference[seq]{ (\Theta, \Delta) \vdash e_1 : \rho_1 ; (\Theta', \Delta') \\ (\Theta', \Delta') \vdash e_2 : \rho_2 ; (\Theta'', \Delta'')}{(\Theta; \Delta) \vdash e_1 ; e_2  : \rho_2 ; (\Theta'', \Delta'')}

\end{center}

\bigskip
%
%

\caption{Big step Operational semantics for the core language}
\label{fig:opsemantics2}
\end{figure*}
\clearpage
\subsection{Results}
\label{subsec:result}
\begin{figure*}[htbp]
\begin{lstlisting}[xleftmargin=\fboxsep, xrightmargin=-\fboxsep, basicstyle=\selectfont\ttfamily\scriptsize, numbers=left, extendedchars=true, numberstyle=\tiny, breaklines=true, mathescape=true] 
state SizedList case of List{
	type SizedListTy : Pi (n) -> List;
	var SizedCons head;
	var SizedList tail;
	method void prepend(elem)[unique SizeListTy(n) -> List >> unique SizedListTy(n+1) -> List]{
		this.head = new SizedCons {var value = elem; var SizedCons next = this.head;};
		this <- SizedListTy(n+1) -> List;
	}
	method void add(elem)[unique SizedListTy(n) -> List >> unique SizedListTy(n+1) -> List]{
		this.tail = new SizedList {var head = new elem; var tail = new plaid.lang.NIL;};
		this <- SizedListTy(n+1) -> List;
	}
	method void append(unique SizedListTy(m) -> List list)[unique SizedListTy(n) -> List >> unique SizedListTy(n+m) -> List]{
		match (list.tail){
			case Nil{
			this.tail = this.tail.add(list.head);	
			this <- SizedList(n+1) -> List;	}
			case Cons{
			this.tail = this.tail.append(new Cons {var value = list.head.value; var next= list.tail;});
			this <- SizedList(n+m) -> List;	}
			default{java.lang.System.out.println("bad");}
		};
	}
	method reverse()[unique SizedListTy (n) -> List >> unique SizedListTy (n) -> List]{
		match (this){
			case Nil{ this;	}
			case Cons{
				new Cons{var value = this.tail.reverse(); var next = this.tail;};
			}
			default{this;}
		};	
	}
}
\end{lstlisting}
\caption{Statically checked SizedList using p-typestate}
\label{fig:sizedList}
\end{figure*}

\begin{figure*}[htbp]
	\begin{lstlisting}[xleftmargin=\fboxsep, xrightmargin=-\fboxsep, basicstyle=\selectfont\ttfamily\scriptsize, numbers=left, extendedchars=true, numberstyle=\tiny, breaklines=true, mathescape=true] 
	state StackModel{
	var Stack k1 = new Stack; // regular stack
	var Stack k2 = new Stack;
	type StackModelType : Pi (c1, c2) -> Stack;
	type StackElementWithId : Pi (b1, b2) -> BoundedInteger;
	// Apply binary pop on both the objects 
	method void push(unique StackElementWithId (b1, b2) -> BoundedInteger element) [unique StackModelType (c1, c2) -> Stack >> unique StackModelType (c1', c2' , t1 = c1 * 2, t2 = b1 * 1, c1' = t1 + t2 , t3 = c2 * 2, t4 = b2 * 1, c2' = t3 + t4) -> Stack]
	{
	this.k1.push(b1);
	this.k2.push(b2);
	this <- StackModelType (c1', c2', c1' = (c1 * 2) + (b1 * 1 ), c2' = (c2 * 2) + (b2 * 1)) -> Stack;
	
	}
	method void pop(unique StackElementWithId (b1, b2) -> BoundedInteger element) [unique StackModelType (c1, c2, b1 == c1, b2 == c2) -> Stack >> unique StackModelType (c1', c2', c1' = c1 / 2, c2' = c2'/2) -> Stack]{
	this.k1.pop();
	this.k2.pop();
	this <- StackModelType (c1', c2', c1' = c1/2 , c2' = c2/2 ) -> Stack;
	}
	\end{lstlisting}
	\caption{Stack simulation using p-typestate}
	\label{fig:stackmodel}
\end{figure*}

\subsection{Type System}
\label{subsec:type-system}


\begin{figure*}[h]
\begin{center}
\hspace*{5ex} \inference[T-Pts-F]{ \Phi , \Gamma \vdash t : \star & \Phi, \Gamma \vdash {\phi : t} & \Phi, \Gamma;(\phi : t) \vdash S : \star \\ t \in \Psi } { \Gamma , \Phi \vdash \Pi({\phi : t, s : S}).\tau : \star } 

\bigskip

\end{center}

\begin{center}
\hspace*{5ex} \inference[T-Pts-I]{ (\Phi, \Gamma) \vdash {\phi : t} & (\Phi, {\phi : t}), \Gamma \vdash s : S    } { (\Phi, \Gamma ) \vdash \textnormal{type $\gamma$} \Pi(\phi : t, s : S).\tau : \Pi({\phi : t, s : S}).\tau } 

\bigskip

\end{center}

\begin{center}
\hspace*{5ex} \inference[T-Pts-C]{ ( \Phi, \Gamma) \vdash \textnormal{type $\gamma$} (\phi : t, s : S).\tau : \Pi({\phi : t, s : S}).\tau \\  (\Phi, \Gamma \vdash \phi_c : t) & (\Phi, (\Gamma; {\phi_c : t})) \vdash s_c : S } {  (\Phi, \Gamma) \vdash [\phi_c / \phi, s_c / s]\tau : (\phi_c, s_c).\tau  } 

\bigskip

\end{center}

\begin{center}
\hspace*{5ex} \inference[T-Pts-C-Eq]{ (\Phi, \Gamma) \vdash \phi_1 = \phi_2 : t & (\Phi, \Gamma) \vdash s_1 = s_2 : S } { (\Phi, \Gamma) \vdash (\phi_1, s_1).\tau  = (\phi_2, s_2).\tau : \star } 

\bigskip

\end{center}

\begin{center}
\hspace*{5ex} \inference[T-Eq]{ (\Phi, \Gamma) \vdash e : \tau_1 & (\Phi, \Gamma) \vdash \tau_1 = \tau_2 : *} { (\Phi, \Gamma) \vdash e : \tau_2 } 

\bigskip

\end{center}

\caption{Type-Family formation, introduction, computation and equality rules}
\label{fig:typeformation}

\end{figure*}

\begin{figure*}[htbp]
\begin{minipage}{0.5\textwidth}
\hspace*{5ex} \inference[T-var]{ (\Phi, \Gamma) \vdash \tau :: * & (x, \tau) \in \Gamma } { (\Phi, \Gamma) \vdash x : (\Phi, \tau) } 

\bigskip
\end{minipage}
\begin{minipage}{0.5\textwidth}
\hspace*{5ex} \inference[T-new]{ decl = \textnormal{state S case of Sup \{...\}} & decl \in ST \\ \tau = (1, S)} {(\Phi, \Gamma) \vdash \textnormal{new S ($\bar{e_1}$)} : (\Phi, \tau)} 
\bigskip
\end{minipage}

\begin{center}
\hspace*{5ex} \inference[T-new-Dep]{decl = \textnormal{state $S_1$ case of Sup \{...\}} & decl \in ST & (\Phi, \Gamma) \vdash (\phi_1, S_1).\tau \ type} {(\Phi, \Gamma) \vdash \textnormal{new $S_1$ ($\phi_{1}$)} : ( \Phi \wedge \phi_1 ,(\phi_1, S_1).\tau) }

\bigskip

\end{center}
\begin{minipage}{0.5\textwidth}
 \begin{center}
\hspace*{5ex} \inference[T-fref]{ (\Phi, \Gamma) \vdash e : (\Phi_1 , \tau_e) \\ \tau_e = (\phi_e, S_e).\tau 
\\ \textnormal{decl = state $S_e$ case of S \{ $\bar{ts}$ ; $\bar{fs}$ ; $\bar{ms}$ \} }  
\\ decl \in ST \\ f \in \bar{fs} \\ ( \Phi_1, (\Gamma, e : \tau_e) \vdash f : ( \Phi_1, \tau)} {(\Phi, \Gamma) \vdash \textnormal{e.f} : (\Phi_1, \tau) }

\bigskip

\end{center}  
\end{minipage}
\hspace{1cm}
\begin{minipage}{0.5\textwidth}
\begin{center}
\hspace*{5ex} \inference{ (\Phi, \Gamma \vdash e_1 : (\Phi_1, \tau_1) \\ ( \Phi_1, (\Gamma, e_1 : \tau_1)) \vdash e : ( \Phi_2, \tau_1) } { (\Phi , \Gamma) \vdash \textnormal{e $\leftarrow$ $e_1$} : (\Phi_2, \tau_1)}[T-update]
 
\bigskip
\end{center}  
\end{minipage}

\begin{center}
\hspace*{5ex} \inference[T-match]{(\Phi, \Gamma) \vdash e_1 : (\Phi_1, \tau_1) & (\Phi_1, (\Gamma, e_1 : \tau_1)) \vdash \overline{case \ e_i : (\Phi_i, \tau_i \rightarrow \tau_{b_i})} \\
 \forall i. \ \tau_i <: \tau_1 & \Phi_u = \bigvee \Phi_i & \tau_u = \bigwedge \tau_{b_i}} 
				  {(\Phi, \Gamma) \vdash \textnormal{match $e_1$  $\overline{case \ e_i}$} : (\Phi_u, \tau_1 \rightarrow \tau_u)}

\bigskip

\end{center}

\begin{center}
\hspace*{5ex} \inference[T-let]{(\Phi, \Gamma) \vdash e_1 : ( \Phi_1, \tau_1) & ( \Phi_1, \Gamma , x : \tau_1 , e_1 : \tau_1) \vdash e : (\Phi_2, \tau)} {(\Phi, \Gamma) \vdash \textnormal{ let x = $e_1$ in e} : ( \Phi_2, \tau)}

\bigskip

\end{center}

\begin{center}
\hspace*{5ex} \inference[T-case]{ (\Phi, \Gamma) \vdash e : (\Phi_1, \tau_1) & (\Phi_1, (\Gamma , e_1 : \tau_1) \vdash e_{b} : ( \Phi_2, \tau_b )} 
{(\Phi, \Gamma) \vdash \textnormal{case  $e$  \{  $e_{b}$ \}} : (\Phi_2, \tau_1 \rightarrow \tau_b) }

\bigskip

\end{center}  

\begin{center}
\hspace*{5ex} \inference[T-mcall]{ (\Phi, \Gamma) \vdash e : (\Phi_1, \tau_b) & \tau_b =  (\phi_b, S_b).\tau \\
 mtype(m , S_b) = \textnormal{$T_r$ m($\overline{T_{i} >> T_{i}'} a _i$)[$\overline{T_{this} >> T_{this}'}$]}\{ e_m \} \\
(\Phi_1, (\Gamma, e : \tau_b) \vdash \tau_b <: T_{this} & (\Phi_1, (\Gamma, e : \tau_b) \vdash \overline{e_i: (\Phi_i, \tau_i)} & \overline{\tau_i <: T_i} 
\\ ((\Phi_1 \wedge (\bigwedge_{i} \Phi_i) (\Gamma, e : \tau_b, \overline{e_i : \tau_i)}) \vdash e_m : (\Phi_r, T_r)} 
{(\Phi,\Gamma) \vdash \textnormal{e.m($e_1, e_2, ... e_p$)} : (\Phi_r, T_r)}

\bigskip

\end{center}  

\begin{center}
\hspace*{5ex} \inference[T-while]{ (\Phi, \Gamma) \vdash e_1 : (\Phi_1, bool) & \Phi_1 \vDash \exists.\phi 
\\ (\Phi_1 \wedge (e_1 == true), (\Gamma, e_1 : bool)) \vdash e : (\Phi_2, \tau) & \Phi_2 \vDash \exists.\phi  \\
(\Phi_1 \wedge \Phi_2 \wedge (e_1 == false) \vDash \exists.\phi } 
{(\Phi, \Gamma) \vdash \textnormal{while [$\exists. \phi$] ($e_1$) \{e\}} : (\Phi_2, \tau) }

\bigskip

\end{center}  

\caption{p-Typestate typing rules for expressions}
\label{fig:expressiontyping}
\end{figure*}



\begin{figure*}[htbp]
\begin{center}
\hspace*{5ex} \inference[T-f Decl ]{ (\Phi, \Gamma) \vdash \textnormal{$\tau$  type} }
{(\Phi, \Gamma) \vdash  \textnormal{ $\tau$ f} : (\Phi, *)  }

\bigskip

\end{center} 

\begin{center}
\hspace*{5ex} \inference[T-m Decl ]{ (\Phi, \Gamma) \vdash e_1 : (\Phi_1, \tau_1) \\
				      (\Phi_1, (\Gamma, e_1 : \tau_1)) \vdash e_2 : (\Phi_2, \tau_2)
				      ... \\
				      (\Phi_{m -1}, (\Gamma, e_1 : \tau_1 ... e_{m-1} : \tau_{m-1})) \vdash e_m : (\Phi_m, \tau_m) \\
				      (\Phi_{m}, (\Gamma, e_1 : \tau_1 ... e_{m} : \tau_{m}, this : \tau_{this})) \vdash e : (\Phi_m, (\Gamma', \tau_r)) \\
				      \forall i. \Gamma'(e_i) = \tau_i' & \Gamma'(this) = \tau_{this}'} 
{(\Phi, \Gamma) \vdash \textnormal{$\tau_r$ m ($\overline{\tau_i >> \tau_i' e_i}$)[$\tau_{this}$ $>>$ $\tau_{this}'$] \{$e$\}} : (\Phi_m, *)}

\bigskip

\end{center}

\begin{center}
\hspace*{5ex} \inference[T-s Decl ]{ \forall f \in fs. (\Phi, \Gamma) \vdash f : (\Phi', * ) \\
\forall t \in tf. (\Phi, \Gamma) \vdash t : (\Phi', * ) \\
\forall m \in ms. (\Phi, \Gamma) \vdash m : (\Phi', * ) \\
(\Phi', \Gamma) \vdash e : (\Phi'' , \tau) } 
{(\Phi, \Gamma) \vdash \textnormal{state S case of S' \{ tf ; fs ; ms ; e \}} : (\Phi'', *)}

\bigskip

\end{center} 

\caption{Formation Rules for Field, Method and State Declarations}
\label{fig:decltyping}
\end{figure*}

\begin{figure*}[h]
\begin{minipage}{.5\textwidth}
\begin{center}
\hspace*{5ex} \inference[T-Sub-Refl]{ \Gamma , \Phi \vdash \tau \ type  } { \tau <: \tau}

\bigskip

\end{center}  
\end{minipage}
\begin{minipage}{.5\textwidth}
\begin{center}
\hspace*{5ex} \inference{ \Gamma , \Phi \vdash \tau_1 <: \tau_2 & \tau_2 <: \tau_3 } {\Gamma , \Phi \vdash \tau_1 <: \tau_3}[T-Sub-Trans]

\bigskip

\end{center}

\end{minipage}

\begin{minipage}{.5\textwidth}

\begin{center}
\hspace*{5ex} \inference[T-Sub-State]{ sdecl \ = \ state \ S \ case \ of \ S_1\{...\} \\ sdecl \in ST } { S <: S_1}
\bigskip

\end{center}  

\end{minipage}
\hspace{1cm}
\begin{minipage}{.5\textwidth}

\begin{center}
\hspace*{5ex} \inference{ \tau_1 = (a_1, \tau_{1'}) & \tau_2 = (a_2, \tau_{2'}) \\ \Gamma , \Phi \vdash a_1  = a_2 & \tau_{1'} <: \tau_{2'}  } { \tau_1 <: \tau_2}[T-Sub-Str]

\bigskip
\end{center}  
\end{minipage}

\begin{minipage}{.5\textwidth}

\begin{center}
\hspace*{5ex} \inference[T-Sub-DepTerm]{ \Phi \vdash \phi_1 \ type , \ \phi_2 \ type \\ \phi_1 \models \phi_2 } { \phi_1 <: \phi_2 }

\bigskip

\end{center}

\end{minipage}
\begin{minipage}{.5\textwidth}
\begin{center}
\hspace*{5ex} \inference{ \Gamma , \Phi \vdash \phi_1 <: \phi_2 \\ \Gamma , \Phi \vdash s_1 <: s_2  } {\Gamma , \Phi \vdash (\phi_1, s_1).\tau <: (\phi_2, s_2).\tau}[T-Pts Sub]

\bigskip

\end{center}  
\end{minipage}
\caption{Subtyping Rules}
\label{fig:subtyping}
\end{figure*}

\end{document}